\documentclass[final,3p,times]{elsarticle}

\usepackage{graphicx}

\usepackage{mathtools,amssymb,amsthm,mathrsfs,flexisym,xfrac}

\usepackage{lineno}

\usepackage{listings}
\usepackage{color}
\usepackage{booktabs}

\usepackage{threeparttable}

\usepackage{multirow}
\usepackage{arydshln}

\usepackage[shortlabels]{enumitem}

\usepackage[colorlinks,linkcolor=blue]{hyperref}
\usepackage[symbol]{footmisc}

\newcounter{bla}

\journal{Computer Physics Communications}

\definecolor{codegreen}{rgb}{0,0.6,0}
\definecolor{codegray}{rgb}{0.5,0.5,0.5}
\definecolor{codepurple}{rgb}{0.58,0,0.82}
\definecolor{backcolor}{rgb}{0.95,0.95,0.92}

\lstdefinestyle{mystyle}{
    numbers=left,
    columns=fixed,
    backgroundcolor=\color{backcolor},   
    commentstyle=\color{codegreen},
    keywordstyle=\color{magenta},
    numberstyle=\tiny\color{codegray},
    stringstyle=\color{codepurple},
    basicstyle={\footnotesize\ttfamily},
    breakatwhitespace=false,         
    breaklines=true,                 
    captionpos=b,                    
    keepspaces=true,                 
    numbersep=5pt,                  
    showspaces=false,                
    showstringspaces=false,
    showtabs=false,                  
    tabsize=4,
}

\lstdefinestyle{bashstyle}{
    columns=fixed,
    backgroundcolor=\color{backcolor},   
    commentstyle=\color{codegreen},
    keywordstyle=\color{magenta},
    numberstyle=\tiny\color{codegray},
    stringstyle=\color{codepurple},
    basicstyle={\small\ttfamily},
    breakatwhitespace=false,         
    breaklines=true,                 
    captionpos=b,                    
    keepspaces=true,                 
    numbersep=5pt,                  
    showspaces=false,                
    showstringspaces=false,
    showtabs=false,                  
    tabsize=4,
}

\setenumerate[1]{leftmargin=*,itemsep=1pt,partopsep=0pt,parsep=0pt,topsep=2pt}
\setitemize[1]{leftmargin=*,itemsep=1pt,partopsep=0pt,parsep=0pt,topsep=2pt}

\begin{document}

\begin{frontmatter}



\title{NNPred: A Predictor Library to Deploy Neural Networks in Computational Fluid Dynamics software}




\author[BUAAadd]{Weishuo Liu}
\ead{liuweishuo@buaa.edu.cn}
\author[JCAdd]{Ziming Song} \ead{zmsong@chiponeic.com}
\author[STFCadd]{Jian Fang\corref{cortextcontent}}
\ead{jian.fang@stfc.ac.uk}\cortext[cortextcontent]{Corresponding author.}
    
\address[BUAAadd]{School of Energy and Power Engineering\unskip, Beihang University\unskip, 37 Xueyuan Road, Haidian District\unskip, Beijing\unskip, 100191\unskip, China}

\address[JCAdd]{Chipone Technology (Beijing) Co., Ltd. \unskip, 2 Jingyuan North Street\unskip, Beijing Economic-Technological Development Area\unskip, Beijing\unskip, 100176\unskip, China}

\address[STFCadd]{Scientific Computing Department\unskip, Science and Technology Facilities Council\unskip, Daresbury Laboratory\unskip, Keckwick Lane\unskip, Daresbury, Warrington\unskip, WA4 4AD\unskip, UK}

\begin{abstract}
A neural-networks predictor library has been developed to deploy machine learning (ML) models into computational fluid dynamics (CFD) codes. The pointer-to-implementation strategy is adopted to isolate the implementation details in order to simplify the implementation to CFD solvers. The library provides simplified model-managing functions by encapsulating the TensorFlow C library, and it maintains self-belonging data containers to deal with data type casting and memory layouts in the input/output (I/O) functions interfacing with CFD solvers. On the language level, the library provides application programming interfaces (APIs) for C++ and Fortran, the two commonly used programming languages in the CFD community. High-level customized modules are developed for two open-source CFD codes, OpenFOAM and CFL3D, written with C++ and Fortran, respectively. The basic usage of the predictor is demonstrated in a simple data-driven heat transfer problem as the first tutorial case. Another tutorial case of modeling the effect of turbulence in channel flow using the library is implemented in both OpenFOAM and CFL3D codes. The developed ML predictor library provides a powerful tool for the deployment of ML models in CFD solvers.
\end{abstract}


\end{frontmatter}



{\bf PROGRAM SUMMARY}

\begin{small}
\noindent
{\em Program Title:} NNPred \\
{\em CPC Library link to program files:} (to be added by Technical Editor) \\
{\em Developer's respository link:} https://github.com/Weishuo93/NN_Pred \\
{\em Code Ocean capsule:} (to be added by Technical Editor)\\
{\em Licensing provisions:}  MIT\footnote[1]{The core predictor for C++ and Fortran is published under MIT license}, LGPL\footnote[2]{The OpenFOAM extension is published under LGPL license}, Apache-2.0\footnote[3]{The CFL3D extension is published under Apache-2.0 license} \\
{\em Programming language:} C/C++, Fortran \\
{\em Nature of problem:}\\
  Coupling machine learning models with CFD programs with minimal modifications in CFD codes. \\
{\em Solution method:}\\
  Adopting pointer-to-implementation (Pimpl) strategy to isolate the implementation details and developing different levels of API and modules on programming language and CFD solvers respectively. \\
{\em Additional comments including restrictions and unusual features :}\\
  Currently rely on TensorFlow as the machine learning model's backends. \\
   \\
\end{small}

\section{Introduction}
In recent years, the machine learning (ML) technology has been empowering the development of computational fluid dynamics (CFD) in many aspects, with specific applications in turbulence modeling~\cite{RN38, RN36}, reduced order modeling~\cite{taira2020modal, parish2020adjoint, SUN2020112732}, uncertainty quantification~\cite{wu2016bayesian, ling2017uncertainty, xiao2019quantification, geneva2019quantifying}, flow control~\cite{gautier2015closed, duriez2017machine}, grid generation~\cite{chedid1996automatic,zhang2020meshingnet, triantafyllidis2002finite}, and etc. ML technology owns its advantage of extracting complex mapping relations from collected data, therefore can speed up the empirical modeling process that would otherwise require a significant amount of human experience. More significantly,  such ML-integrated applications generally have superior performance against empirical methods especially for complex systems (e.g., turbulence). Numerous elaborate revisions have demonstrated the feasibility of integrating ML techniques in a CFD toolset~\cite{brenner2019perspective, Brunton2020machine}, indicating their huge potential in both academic and industrial scenarios.  

However, there exists a noticeable barrier in terms of the programming infrastructures between typical ML workflow and conventional CFD solvers, since the mainstream programming languages are fundamentally different between the two communities.
For training the ML models, interpretive languages (e.g., Python) are generally adopted for their convenience in run-time execution and searching for optimal configuration of the model~\cite{RN100, RN84, paszke2019pytorch, RN68}. But for CFD solvers, compiled languages such as C/C++ and Fortran are widely used for the efficiency of massive floating-point operations~\cite{jasak2007openfoam, krist1998cfl3d, economon2016su2, archambeau2004code}. Researchers or engineers might need to spend a great amount of time and effort implementing and testing an ML model within a CFD solver.

From the technical perspective, deploying a trained ML model to an application program can take the following options. The first option is to drive the Python interpreter~\cite{van1995extending} inside the CFD program. But at each iteration step, the in-memory data needs to be packed as Python objects and passed to the model prediction function written in Python, and vice versa for the returned Python arrays~\cite{RN55, RN41}. The implementation can be highly customized, but the massive converting process would cause a significant overhead of computer resources, making it only suitable for some small-scale calculations. Another method is to reconstruct the ML model with the CFD's native language according to the model's information saved from Python scripts. This strategy could maintain good compatibility with the CFD program, and has been adopted by the Fortran-Keras Bridge library~\cite{ott2020fortran}. However, active development is needed to keep up with the updates of ML libraries, which are currently fast-changing, and if one were to reconstruct a complex NN architecture in CFD software, some new features from the updated ML library might be not available timely. However, most wide-used ML frameworks (e.g., TensorFlow and PyTorch) provide binary libraries with native binding application programming interfaces (APIs) in fundamental languages (e.g., C/C++), which might be compatible with CFD codes. Their computational efficiency is normally optimized internally, and the newest features are naturally updated. Therefore, utilizing the ML libraries' binding APIs became the best choice to take into account the computational efficiency, complex model structures, and coding compatibility. For example, Geneva and Zabaras deployed a neural network (NN) from PyTorch~\cite{paszke2019pytorch} into OpenFOAM~\cite{jasak2007openfoam} for uncertainty quantification studies~\cite{geneva2019quantifying}. Maulik~\textit{et~al.}~\cite{maulik2021deploying} employed the TensorFlow (TF) C-API~\cite{TFCWeb} also in OpenFOAM and formulate Reynolds-Averaged Navier--Stokes (RANS) and large eddy simulation (LES) solvers calling runtime ML predictions. Shin~\textit{et~al.} modeled subgrid-scale closure in OpenFOAM LES solver via OpenVINO~\cite{shin2021data}, which is an artificial intelligence (AI) inference toolkit designed for Intel CPU.

Nonetheless, the deployment of ML models still remains a challenge for CFD programmers. It may take developers a lot of effort to find the APIs to call a model's forward prediction. Besides, since ML models are normally trained on GPU but CFD codes are mainly applied on CPU clusters, problems might exist in data type casting and parallelization. To deal with the above-mentioned difficulties, we develop a neural-networks (NN) predictor library under the pointer-to-implementation (Pimpl) strategy~\cite{RN64, gamma1995design}, which isolates the implementation detail so that its implementation to CFD software can be largely simplified. The library provides simplified model-managing functions by encapsulating the TensorFlow C-API, and maintains self-belonging data containers to automatically deal with data type casting and memory layouts in the I/O functions interfacing with CFD solvers. The process of the implementation is simplified to two major steps: initialization and prediction, where each step can be achieved using only 3-4 lines of scripts. The library currently supports APIs for two languages: C++ and Fortran, with high-level customized modules for two open-source CFD codes, OpenFOAM (OF) and CFL3D~\cite{krist1998cfl3d}. The former is written in C++, following the object-oriented programming (OOP) paradigm, whereas the latter is NASA's compressible CFD solver written in Fortran, and has been widely applied in many projects~\cite{rumsey1997cfl3d} since the 1980s. 

In this paper, two tutorial cases are presented to demonstrate the usage of the library. First, a heat transfer problem is considered to introduce the basic process to call the ML model's prediction from a general partial differential equation solver. The Laplacian solver of the OpenFOAM is modified for this application. In the second tutorial case, the methodology of the iterative ML-RANS framework of Liu \textit{et al.}~\cite{liu2021an, liu2022on} is reproduced in both OpenFOAM and CFL3D. The effect of turbulence in a channel flow is modeled by NN in this case. The computational efficiency and parallel performance are also tested for both codes. By maintaining an independent predictor in each thread to perform serial local predictions, the CFD program's original message passing interface (MPI) pattern~\cite{gropp1999using} does not need to be re-designed, and no extra message passing load is increased. The source code repository is publicly available and free of use\footnote[4]{\url{https://github.com/Weishuo93/NN_Pred}}.

The paper is organized as follows. Section~\ref{Sec:Description} describes the program design of the library, including its basic application workflow, API hierarchy, and a minimal example explaining the runtime model controlling and data I/O mechanism. In Sec.~\ref{Sec:Cases}, the tutorial cases, i.e., the heat transfer and turbulent channel flow, are presented. Conclusions are drawn in the last section (Sec.~\ref{Sec:Conclusion}).

\section{Library Design\label{Sec:Description}}
In this section, the program design of the predictor library is introduced. First, the library's typical application workflow with its fundamental structure is shown in subsection.~\ref{SubSec:Structure} to provide an initial intuitive understanding of where and how this library should be used. Then the Pimpl concept is introduced to develop a well-isolated pure-method class, based on which the API hierarchy in Fortran and higher-level CFD programs are established (subsection.~\ref{SubSec:Pimpl}). Finally, the predictor's model loading and runtime data I/O mechanisms are demonstrated by running a simple \textbf{A + B} model with a minimal C++ code example (subsection.~\ref{SubSec:APIs}). This example is also used in the API manual, documented in \ref{Append:APIs}, to facilitate understanding the API usage in different languages and CFD programs.

\subsection{Library Structure and Basic Workflow\label{SubSec:Structure}}
\begin{figure}[!ht]
  \centering
  \includegraphics[width=1\textwidth,trim=30 155 10 70,clip]{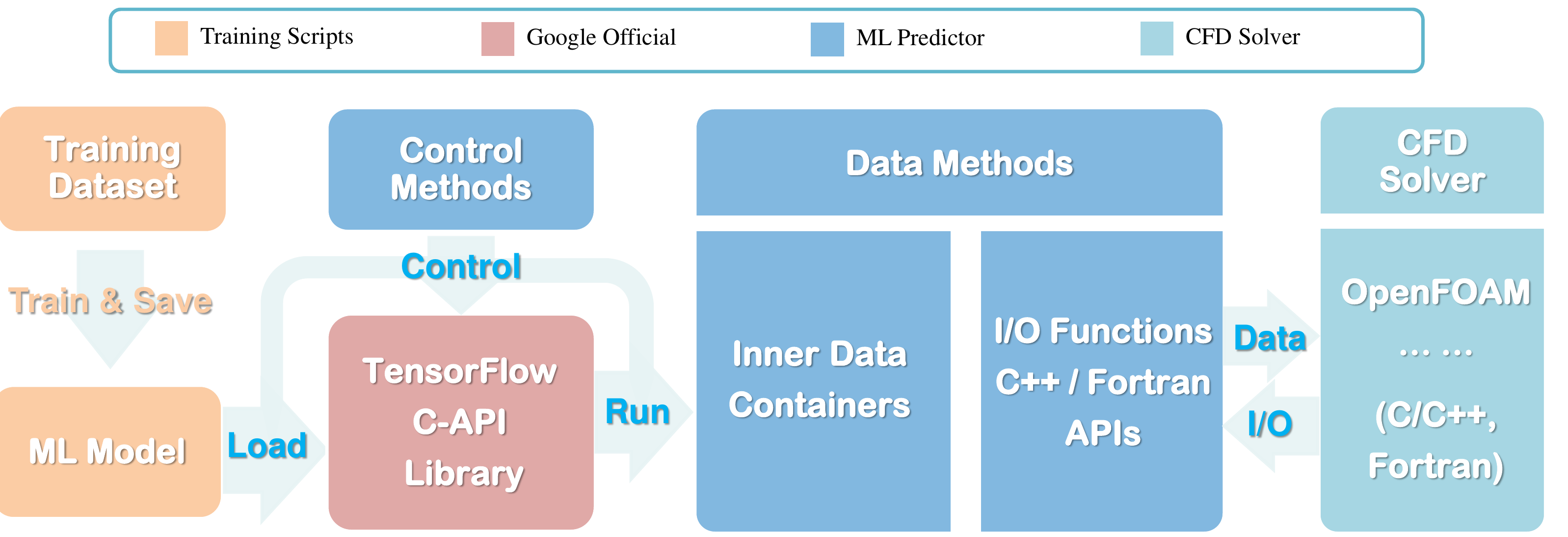}
  \caption{\label{fig:structure}The structure and work flow of the ML predictor library.}
\end{figure}

The typical workflow and the library structure are shown in Fig.~\ref{fig:structure}. The core functionality of this library includes, loading a trained model, performing prediction, and I/O with external data sources. Structurally, this library has two main parts. The first part is the model-control methods, which wraps the TensorFlow C library~\cite{TFCWeb} to impose controls on the model's initialization and prediction. The control methods significantly simplify the calling process, thus reducing the difficulty in deploying the ML model. The other is the data 
methods with self-maintained data containers and the I/O functions from/to external data sources. Normally, the external sources might be from another programming language (e.g., Fortran in many legacy codes) or under a specific data structure (e.g., the \texttt{volScalarField} in OpenFOAM), therefore, 
customized I/O function with data type and memory layout changing options are provided for adaption to CFD codes.

The predictor library is designed with the following features to facilitate the deployment of a ML model:
\begin{itemize}
  \item Supported format\footnote[5]{Explanations of TensorFlow format can be found in Sec.~\ref{SubSec:Formats}}.:
  \begin{itemize}
    \item PB graph (normally saved by TensorFlow version 1 scripts)
    \item SavedModel (normally saved by Keras APIs)  
  \end{itemize}
  \item Parallelization:
  \begin{itemize}
    \item Automatic parallelization on idle CPUs (default setting for TensorFlow C library)
    \item Manually set core numbers for model prediction to adapt a parallel computation
  \end{itemize}
  \item Data I/O:
  \begin{itemize}
    \item Automatic data type casting according to the external source/target arrays
    \item Memory layout changing options (adaption of row/column-major layout)
  \end{itemize}
\end{itemize}

\subsection{The Pimpl Paradigm and API hierarchy \label{SubSec:Pimpl}}

\begin{figure}[!ht]
  \centering
  \includegraphics[width=1\textwidth,trim=20 110 45 100,clip]{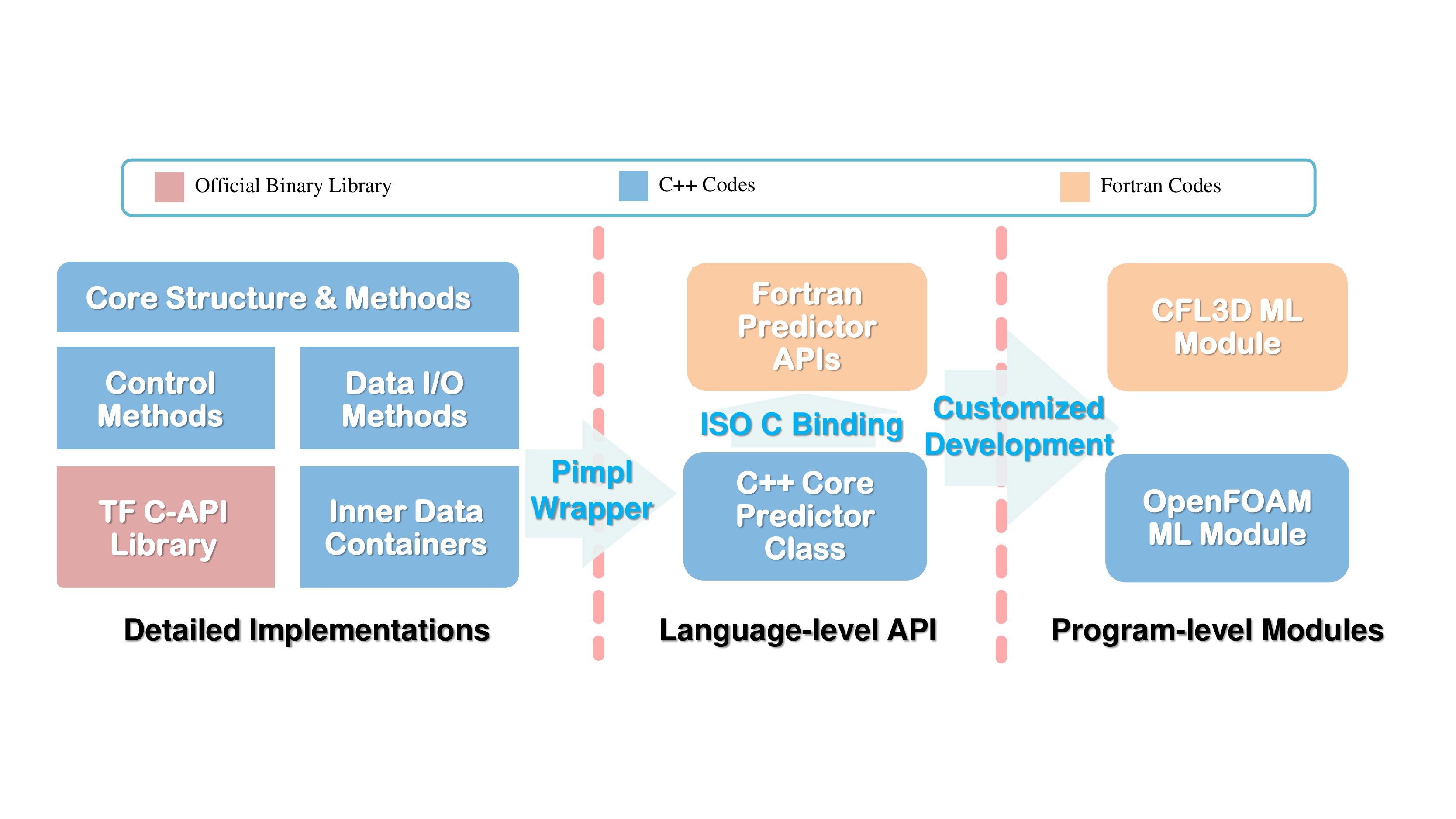}
  \caption{\label{fig:hierarchy}The API hierarchy relationships of the predictor library.}
\end{figure}

The top-level design of this predictor library follows the Pimpl paradigm. It is an implementation-hiding technique in which a public class wraps a structure or class that cannot be seen outside the library~\cite{gamma1995design}. In the predictor library, control methods, data methods, and the inner data container are all well-isolated, and the exposed public class is designed as a pure method class. This wrapping style reconciles the possibility of implementing the library in both object-oriented and process-oriented programs. The API hierarchy is shown in Fig.~\ref{fig:hierarchy}, where the core predictor class is written with C++ under the Pimpl paradigm. The Fortran APIs are then developed via the \texttt{ISO\_C\_BINDING} module~\cite{bush2007new, reid2007new} from the C++ core predictor. The language-level APIs are then further encapsulated into program-level modules for OpenFOAM and CFL3D, which could read settings from text setup files. In this way, the model deployment is further simplified and the modification of CFD codes could be minimized. 

The Pimpl strategy is achieved in C++ with the d-pointer technique~\cite{eckel2000thinking}, which means the public class only needs to maintain a private pointer to the implementation class. The detailed methods and data are actually coded in the implementation class. A demonstrative bare-bone header for the public class is shown in Listing~\ref{lst:Pimpl}, where the implementation class only needs to be declared (line 2) without any further definition, and the public class can utilize the private implementation-class pointer (defined in line 20) to realize all the public methods.

\begin{lstlisting}[language=C++,style=mystyle,caption=Main structure of the predictor library's header, label={lst:Pimpl}]
// Declaration of the Impl class
class PredictorImpl;

// The public class exposed to users
class Predictor {

  public:
    // Constructors:
    explicit Predictor(std::string pbfile);
    // ... ... Other constructors will be introduced in the API section.

    // Destructor:
    virtual ~Predictor();

    // Class methods: initialization, data I/O, model running functions ...
    // ... ... ...    will be introduced in detail in the API section.
    
  private:
    // d-pointer to hide implementation details from the users
    PredictorImpl *d;

};
\end{lstlisting}

\subsection{Minimal example with A+B Model\label{SubSec:APIs}}

We use a simple \textbf{A + B} model to illustrate the basic process of deploying an ML model in C++ code, which includes initializing the model, performing prediction, and interfacing with outer memory spaces. 
This minimal example can be seen as an entry-level getting-started tutorial, and the entire detailed API documentation is also demonstrated with this example, which is documented in~\ref{Append:APIs}. The installation guide with the necessary dependencies and compile options can be found in \ref{Append:Compile}.

\subsubsection{Explanation of the \textbf{A + B} model and its saved formats\label{SubSec:Formats}}
Basically, the predictor library adopts TensorFlow C library~\cite{TFCWeb} to manage models, and therefore, any TensorFlow supported formats are accepted in the present library. At the fundamental level, TensorFlow defines an ML model as sequential records of how data is manipulated mathematically till the outputs are obtained. Each mathematical operation or the input/output data placeholder is recorded as a \textbf{node}, and a series of nodes in a collection constitutes a \textbf{graph}. At higher levels, TensorFlow integrates \textbf{Keras}~\cite{joseph2021keras} to compose models. \textbf{Keras} provides frequently used NN \textbf{layer} APIs by grouping basic \textbf{node} operations, and the model management methods (e.g., training, evaluation, prediction, and saving) are also well encapsulated. The high-level model (i.e., created by \textbf{Keras}) is normally saved in the official \textbf{SavedModel} format. 

\begin{figure}[!ht]
  \centering
  \includegraphics[width=1\textwidth,trim=80 182 80 130,clip]{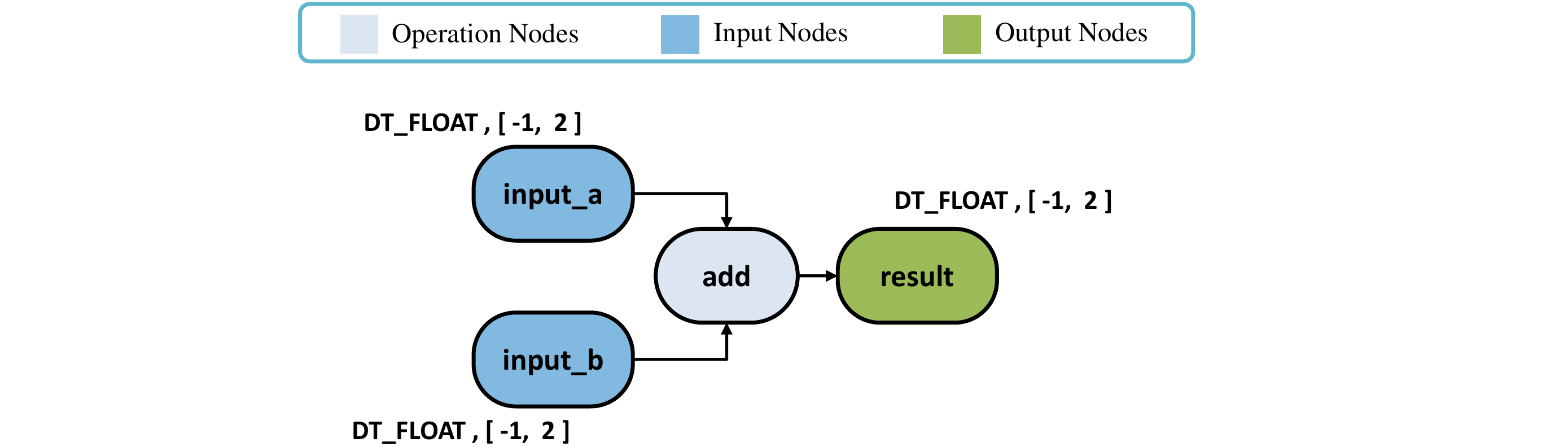}
  \caption{\label{fig:pbgraph}The API hierarchy relationships of the predictor library.}
\end{figure}

The \textbf{A + B} model is simply to get the summation of two arrays, in which the operation of adding two 2D arrays (i.e., \textbf{A} and \textbf{B}) is recorded. The \textbf{A + B} model has two input arrays with dimension size: $n$ $\times$ 2, where n is the number of data instances. The shape of the output is also an $n$ $\times$ 2 array, holding the summation of the two input arrays. The node definitions are shown in Fig.~\ref{fig:pbgraph}, with the data type being specified as the single precision floating point (\texttt{DT\_FLOAT} under TensorFlow's definition). The model can be created in two different ways and stored in the corresponding formats. One way acts bottom-to-top to establish graphs from basic operations, such graphs are saved in protocol buffers (PB) format~\cite{PBWeb} (with file extensions \texttt{*.pb}) as a sequential stream of bytes. While the other way uses \textbf{Keras} to build and saves the models in the \texttt{SavedModel} format (which is actually a folder). The two methods for building and saving ML models can be viewed in the Python script: \href{https://github.com/Weishuo93/NN_Pred/blob/master/Predictor-Core/test/models/createmodel_AplusB.py}{\texttt{createmodel\_AplusB.py}} in the repository. The command-line usage to create the two different formats is shown in ~\ref{Append:APIs}.

Note that the library is upward compatible with TensorFlow. The higher version of TensorFlow can load the model from the library using a lower version of TensorFlow, but the reverse operation would raise an incompatibility problem. The model saved in a TensorFlow version 2 Keras \texttt{SavedModel} format can not be loaded via a TensorFlow version 1 binary library.

\subsubsection{Runtime model control and data I/O via C++ example\label{Sec:CppPredictor}}
In this part, the \textbf{A + B} model is loaded in a C++ code example to demonstrate the runtime model control and data I/O mechanism of the predictor. The PB graph format, \texttt{simple\_graph\_tf2.pb}, is used in the code snippet example, and the alternative APIs for other TensorFlow format is introduced in the documentation (\ref{Append:APIs}).

\begin{table}[htbp!]
  \caption{ Lines of scripts required for each step of the program \label{Table:ProgramStep}}
  \centering
  \begin{threeparttable}
  \begin{tabular*}{\textwidth}{@{\hspace{2em}}l @{\extracolsep{\fill}}clc@{\hspace{2em}}}
    \toprule 
    \multicolumn{2}{c}{\textbf{Initialization}} & \multicolumn{2}{c}{\textbf{Prediction}} \\ \cline{1-2} \cline{3-4}\\[-11pt]
    Program Step              & Lines of scripts \tnote{*}  & Program Step            & Lines of Codes \tnote{*} \\  \\[-11pt]\hline
    Load Model                & 1               & Set Input Data          & $\boldsymbol{N}_{i}$   \\
    Register Nodes            & $\boldsymbol{N}_{io}$   & Run Model      & 1             \\
    Set Data Number           & 1               & Get Output Data         & $\boldsymbol{N}_{o}$    \\
    \bottomrule  
   \end{tabular*}
   \begin{tablenotes}
    \footnotesize
    \item[*] $\boldsymbol{N}_{i}$, $\boldsymbol{N}_{o}$, and $\boldsymbol{N}_{io}$ stands for the number of input nodes, output nodes and their summation. 
    \end{tablenotes}
  \end{threeparttable}
\end{table}

\begin{figure}[!ht]
  \centering
  \includegraphics[width=1\textwidth,trim=40 50 95 40,clip]{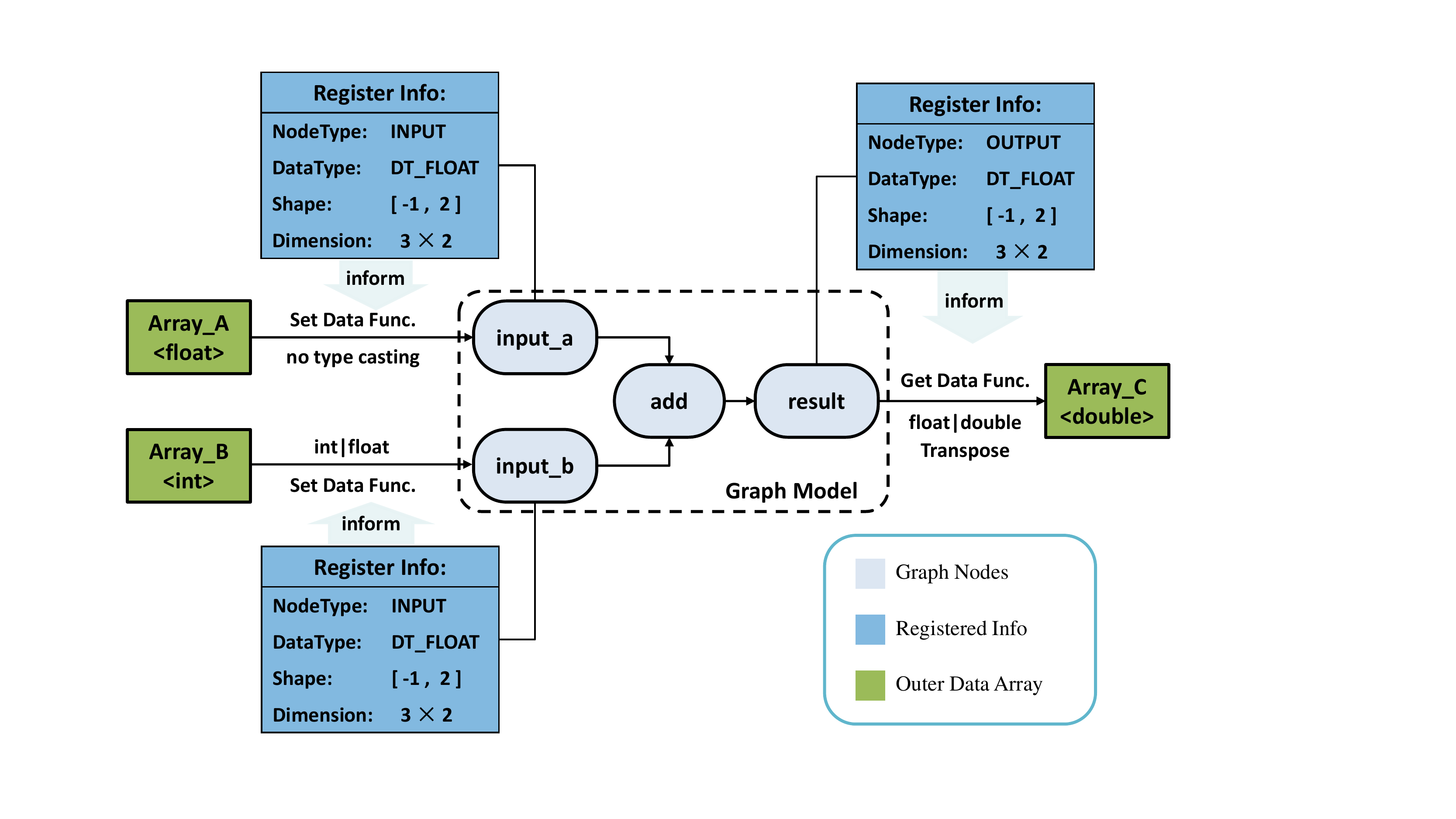}
  \caption{\label{fig:AplusB}Interaction relation within predictor elements and with outer memory spaces during a prediction process.}
\end{figure}

Basically, the overall process can be divided into two major parts: initialization and prediction. Each part can be implemented with several lines of scripts (see Table~\ref{Table:ProgramStep}). The initialization part consists of loading the model, registering the input/output nodes, and setting the number of data instances, whereas the prediction part includes setting the input data, running predictions, and extracting the result.

Figure~\ref{fig:AplusB} illustrates how the different elements in the predictor interact with others, and an example of the scripts is shown in Listing~\ref{lst:AplusB}. The loading procedure is achieved by creating the predictor object with the model's path being specified (line 7 in the code example), in which, all the definitions of nodes are cached (circled by the dotted line in Fig.~\ref{fig:AplusB}). Then users should register nodes by specifying the nodes' name and input/output type (line 11-16 in Listing~\ref{lst:AplusB}). In this process, the pieces of information of the registered nodes are extracted and recorded for the automatic I/O functions (blue tables in Fig.\ref{fig:AplusB}). Finally, the initialization process can be considered completed after setting the number of data instances (line 19 in Listing~\ref{lst:AplusB}), after which the predictor will allocate memory space for the inner data containers according to the recorded data types and dimensions.

\begin{figure}[!ht]
  \centering
  \includegraphics[width=1\textwidth,trim=0 378 0 40,clip]{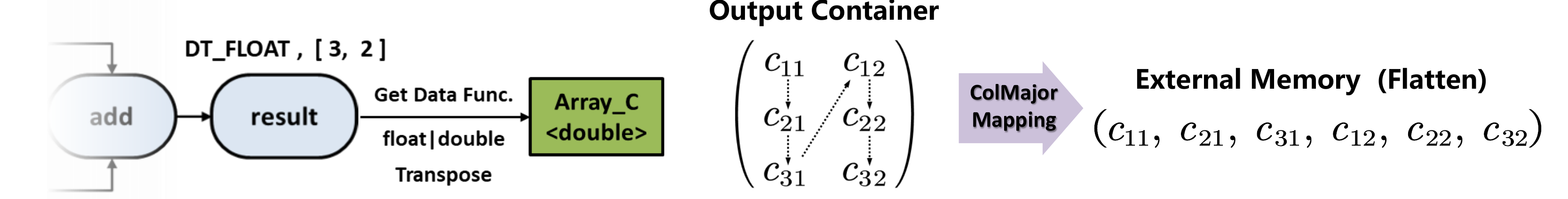}
  \caption{\label{fig:Layout}Extract data and map into a column-majored memory space.}
\end{figure}

Afterward, three C++ arrays are created (line 21-26 in Listing~\ref{lst:AplusB}) to represent the external sources of data. In the prediction process, the input data is first mapped to the predictor's input container through the set data function (line 28-30 in Listing~\ref{lst:AplusB}). Then the model is executed (line 33 in Listing~\ref{lst:AplusB}) and the output containers are filled with computation results, which will finally be extracted to the external arrays via the get-data function (line 36 in Listing~\ref{lst:AplusB}). It needs to be mentioned that the data type of the external arrays is not necessarily consistent with the model-defined type (e.g., float). The set/get data function can automatically cast the data according to the recorded node information (done by the register function) to ensure a correct computational result. For example, the set-data function casts the integers from \texttt{Array\_B} to floats as defined by the node container \texttt{input\_b}. Apart from this, the external array's memory layout can also be specified during the I/O process. Fig.~\ref{fig:Layout} shows the process to extract the output data into a column-majored memory space.

It needs to be mentioned that the actual information passed to the set/get data functions is a pointer that specifies the data type (e.g., \texttt{int}, \texttt{float} or \texttt{double}), the array's memory address, and the number of data indicating the size of this piece of memory. A more detailed explanation including the intrinsic version of setting and getting data functions can be found in the API document~(\ref{Append:APIs}).

\begin{lstlisting}[language=C++,style=mystyle,caption=Minimal example to run A + B model in C++, label={lst:AplusB}]
// Header files
#include "predictor.h"   // Predictor header
#include <vector>        // C++ standard header

int main(int argc, char const *argv[]) {
    // Load Model:
    Predictor pd("simple_graph_tf2.pb"); // Model's path or filename

    // Register node:
    // Inputs: 
    // Predictor::INPUT_NODE is the node type enumerate 
    pd.regist_node("input_a", Predictor::INPUT_NODE);
    pd.regist_node("input_b", Predictor::INPUT_NODE);
    // Outputs:
    // Predictor::OUTPUT_NODE is the node type enumerate 
    pd.regist_node("result", Predictor::OUTPUT_NODE);

    // Set the number of data instances (n=3)
    pd.set_data_count(3);

    // Create external source of input/output data array:
    // Inputs:
    std::vector<float> vec_input1_float = {1.1, 2.2, 3.3, 4.4, 5.5, 6.6};
    std::vector<int> vec_input2_float = {6, 5, 4, 3, 2, 1};
    // Outputs:
    std::vector<double> vec_out_float(6);

    // Set data for input nodes
    pd.set_node_data("input_a", vec_input1);
    pd.set_node_data("input_b", vec_input2);

    // Run model
    pd.run();

    // Get output into the target container
    pd.get_node_data("result", vec_out, Predictor::ColumnMajor);
    
    // Check results, expected calculation results:
    // vec_out: [7.1, 7.3, 7.5, 7.2, 7.4, 7.6]
    //          [C11, C21, C31, C12, C22, C32]    
    return 0;
}
\end{lstlisting}

\section{Tutorial Cases\label{Sec:Cases}}
Two tutorial cases are presented in this section. The first case is to solve a simple heat transfer problem using a data-driven approach, which demonstrates the basic process to invoke an ML model's prediction within a CFD program by modifying the Laplacian solver in OpenFOAM. The second tutorial replicates the methodology of the iterative ML-RANS framework~\cite{liu2021an, liu2022on} in OpenFOAM, and further migrates the turbulence model to the CFL3D code. The simulation cross validates the feasibility of the ML-RANS system, as both CFD codes have achieved converged solutions identical to the high-fidelity database of turbulent channel flow.

\subsection{A Heat Transfer Problem\label{Sec:HeatTransfer}}

The predictor library is first applied in a modified OpenFOAM Laplacian solver (source code can be found in the folder: \href{https://github.com/Weishuo93/NN_Pred/tree/master/OpenFOAM-Extension/tutorials/heat_transfer}{\texttt{heat\_transfer}}) to solve a one-dimensional steady heat transfer problem with radioactive and convective heat sources~\cite{RN41}, which can be defined by the following equation:
\begin{equation}
  \label{eq:heateqn}
  \frac{d^{2} T}{d x^{2}}=\varepsilon(T)\left(T_{\infty}^{4}-T^{4}\right)+h\left(T_{\infty}-T\right),
\end{equation}
in which, $T$ is the temperature, $x$ is the space coordinate with range: $x \in[0, 1]$, $T_{\infty}$ is the temperature of the heating source, $h$ is a constant convection coefficient: $h=0.5$, and $\varepsilon(T)$ is the emissivity determined by the local temperature:
\begin{equation}
  \label{eq:emissivity_theo}
  \varepsilon(T)=\left[1+5 \sin \left(\frac{3 \pi}{200} T\right)+\exp (0.02 T)\right] \times 10^{-4}
\end{equation}

\begin{figure}[!ht]
  \centering
  \includegraphics[width=0.89\textwidth,trim=-80 2 -70 30,clip]{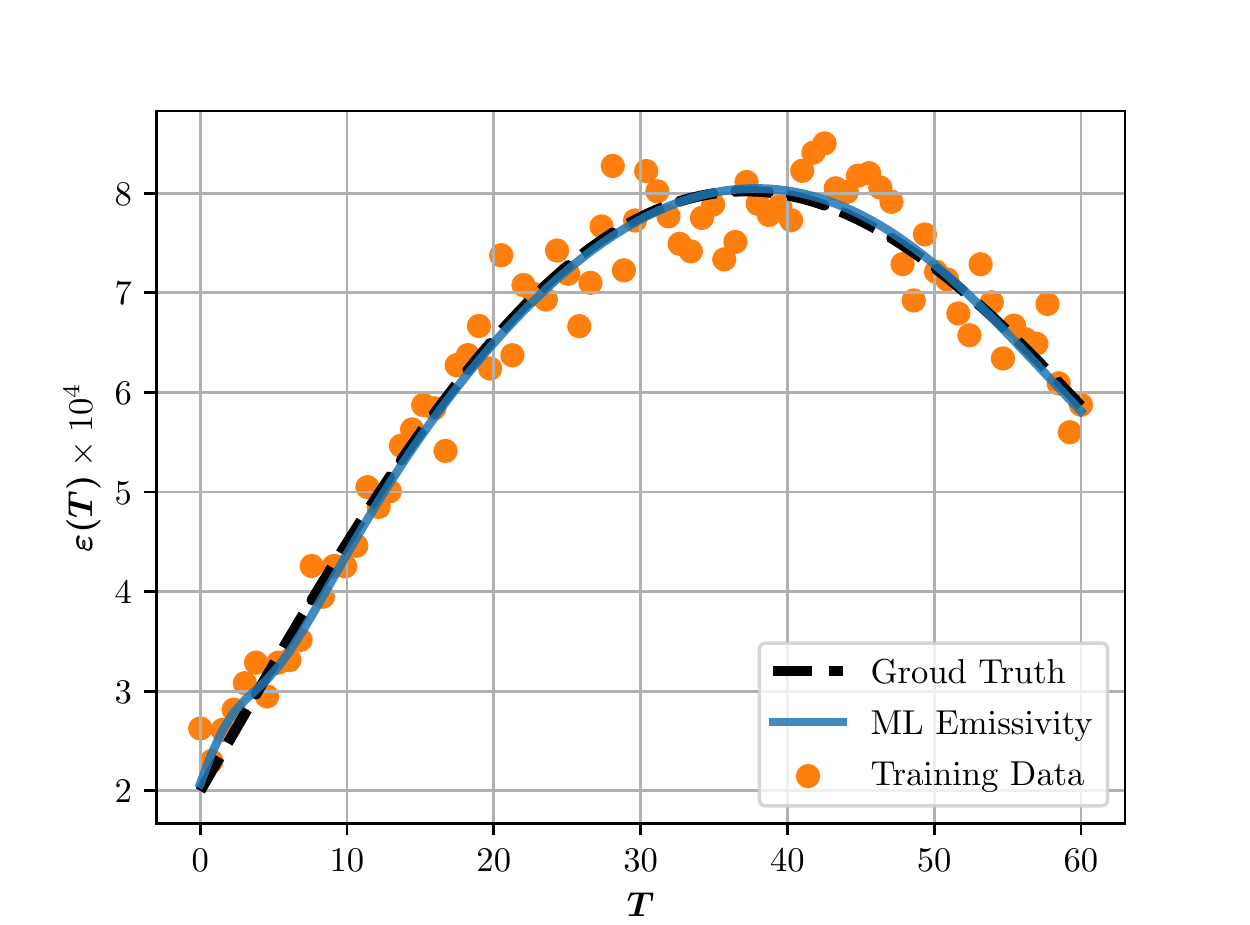}
  \caption{\label{fig:ThermoData_ML} Comparison of machine-learned function, the true distribution and data with noises.}
\end{figure}

However, the emissivity-temperature relation described by Eq.~\ref{eq:emissivity_theo} is supposed to be unknown to the solver, and a model is therefore needed to establish the relation between emissivity and temperature. In the present case, the training data is generated by adding a zero-mean Gaussian noise, $\mathcal{N}\left(0,\sigma^{2}\right)$, to Eq.~\ref{eq:emissivity_theo} ($\sigma$ is taken to be 0.3). The emissivity analytical formula and the training data can be seen in Fig.~\ref{fig:ThermoData_ML}.

To solve Eq.~\ref{eq:heateqn}, a traditional modeling strategy often attempts to find out an empirical analytical relation with coefficients regressed from the observed data, so that the relation can be solved as a part of the governing equations. For this tutorial, a data-driven approach is adopted in the modeling process, where a two-layer NN is trained to approximate the emissivity-temperature relation. The temperature is taken as the input, and the emissivity is the output. The hyper-parameters for the NN training are listed in Table.~\ref{Table:ThermoPara}. The NN-estimated emissivity model compared to the theoretical truth and the training data is shown in Fig.~\ref{fig:ThermoData_ML}.

\begin{table}[htbp!]
  \caption{ Hyper-parameters of emissivity-temperature NN \label{Table:ThermoPara}}
  \centering
  \begin{tabular*}{\textwidth}
      {@{\hspace{6em}}c @{\extracolsep{\fill}}c @{\hspace{6em}}}
  \\[-11pt]
  \toprule 
  \textbf{Hyperparameters}     & \textbf{Recommend Values}  \\[1pt] \hline \\[-9pt]
  Number of Hidden Layers       & 2                         \\
  Number of Nodes per Layer     & 32                        \\
  Activation Function           & tanh                      \\
  Optimizer                     & Adam                      \\
  Learning Rate                 & 0.001                     \\
  L1 Regularization             & 3e-05                     \\
  L2 Regularization             & 8e-04                     \\
  Epoch Numbers                 & 5000                      \\
  \bottomrule  
  \end{tabular*}
\end{table}

\begin{figure}[!ht]
  \centering
  \includegraphics[width=0.88\textwidth,trim=-80 2 -70 10,clip]{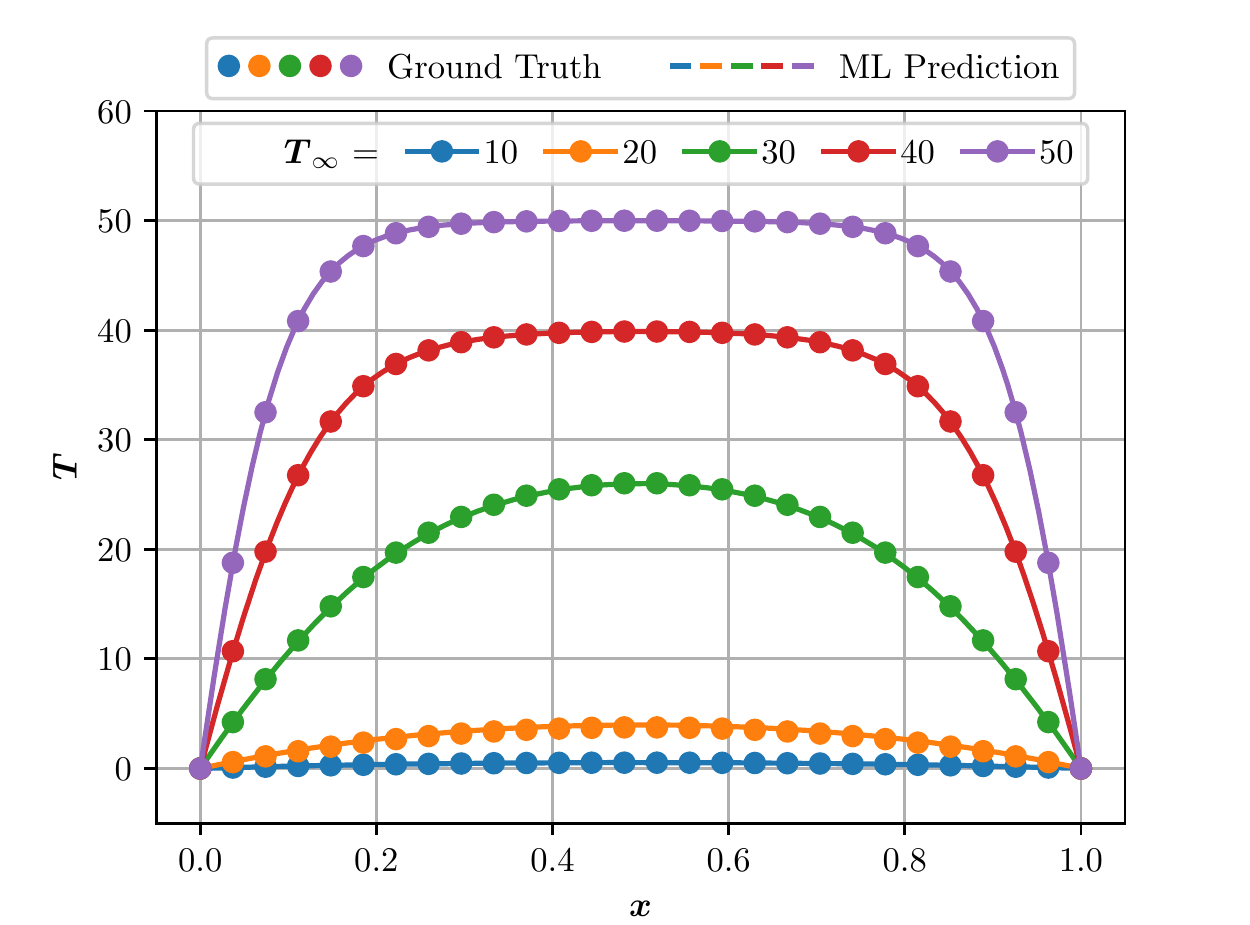}
  \caption{\label{fig:ThermoSim} Temperature distribution of the true analytical model and ML model.}
\end{figure}

The trained model is then deployed in OpenFOAM to solve Eq.~\ref{eq:heateqn} and get the temperature distribution, using the modified Laplacian solver of the OpenFOAM. The comparison of the temperature between the ML and the analytical models is shown in Fig.~\ref{fig:ThermoSim}, where the temperature of the heat source is set to be $T_\infty=10, 20, ... , 50$. As can be seen from the figure, the ML model achieves a remarkable accuracy, indicating the ML model successfully learned the relation described by Eq.~\ref{eq:emissivity_theo} via the training process, and the deployment of the model to the OpenFOAM effectively solves the heat transfer problem. This training and deployment process can be applied to solve a real engineering problem, where there is no analytical formula to describe the unknown relation and only some measured data are available.

\subsection{The ML-RANS model in OpenFOAM and CFL3D\label{Sec:ML-RANS}}

The training methodology of the iterative ML-RANS framework~\cite{liu2021an, liu2022on} is replicated, in which, NN is adopted to model the effect of turbulence in channel flows. The NN model is then deployed through the predictor library in two classic CFD codes, OpenFOAM and CFL3D, written with C++ and Fortran, respectively. The relevant codes including the training scripts, modified solvers and simulation cases can be viewed in the repository (see folder: \href{https://github.com/Weishuo93/NN_Pred/tree/master/OpenFOAM-Extension/tutorials/turbulent_channel}{\texttt{turbulent\_channel}} and \href{https://github.com/Weishuo93/NN_Pred/tree/master/CFL3D-Extension}{\texttt{CFL3D-Extension}} for details).

\begin{figure}[!ht]
  \centering
  \includegraphics[width=0.88\textwidth,trim=120 115 120 120,clip]{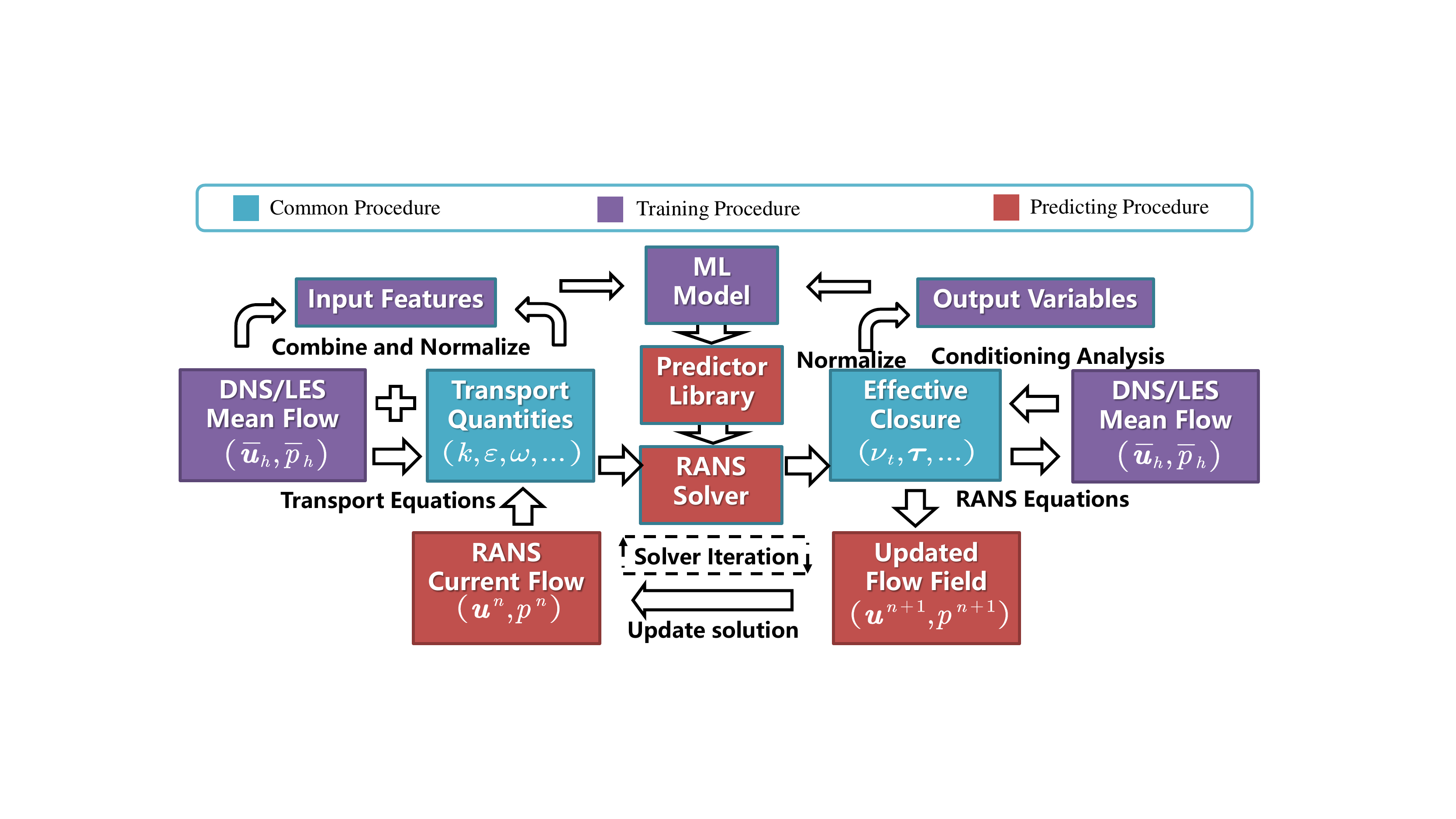}
  \caption{\label{fig:ML_RANS_Frame} The iterative ML-RANS framework with consistent computational procedure.}
\end{figure}

The ML-RANS framework is designed with a built-in consistency to reproduce the high-fidelity training data in the \textit{a posteriori} simulations. The model has been tested and assessed in engineering applications, and it showed an improved performance against conventional models~\cite{fang2022data}. The computational procedures are shown in Fig.~\ref{fig:ML_RANS_Frame}, where the idea is to incorporate transport equations of a conventional turbulence model to provide auxiliary quantities for both training and prediction. Such quantities are combined with mean flow variables to form the dimensionless input features for the ML model. Meanwhile, the form of the closure term is chosen based on the conditioning analysis~\cite{RN50} of the Reynolds-averaged Navier--Stokes (RANS) equations, and the actual effect of the closure term is verified by forward solving the RANS equations based on the closure's pre-mapped frozen distribution. With a proper normalization of the input and output data, the ML model can be trained and then loaded by CFD solvers to perform the \textit{a posteriori} simulations.

The DNS channel flow at $Re_\tau=\left\{180, 395, 640, 1020\right\}$ from Abe \textit{et al.}~\cite{RN62, abe2004surface} are used as training cases, in which, the Reynolds number, $Re_\tau$, is defined as:
\begin{equation}
  Re_\tau=\frac{u_\tau h}{\nu}
\end{equation}
where $u_\tau$ is the friction velocity, $h$ is the half height of the channel, and $\nu$ is the molecular kinematic viscosity.

For each independent training case, the input and output data are obtained with the following steps. First, the DNS mean flow fields (i.e., velocity and pressure fields for incompressible flow) are interpolated to the computational mesh. The transport equations of the $k$--$\omega$ SST~\cite{RN98} model are then solved based on the frozen DNS mean field through the modified scalar transport solver in OpenFOAM, with turbulence kinetic energy $k$ and turbulence frequency $\omega$ being saved as auxiliary variables. 
Second, the eddy-viscosity field is calculated from DNS data as the effective closure term based on the conditioning analysis~\cite{RN50}.
Finally, the saved $k$ and $\omega$ are adopted to perform a bounded normalization~\cite{liu2022on} of the input and output data, with their physical meanings and non-dimensional denominators listed in Table.~\ref{Table:ChannelFeatures}. 
After generating training data for each case, the data from the four training cases are grouped together to train a four-layer NN (see Table.~\ref{Table:RANSPara} for the training parameters). The eddy-viscosity used in the second step is calculated from the mean strain and Reynolds stress tensors as,
\begin{equation}
  \label{eq:eddy_viscosity}
  \nu_{t} = \left| \frac{\tau_{ij}S_{ij}}{2S_{ij}S_{ij}} \right|,
\end{equation}
where $\tau_{ij}$ is the Reynolds stress tensor, $S_{ij}$ is the strain rate tensor, and the Einstein summation is applied for repeated indices. 

\begin{table} [htbp]
\caption{\label{Table:ChannelFeatures}Non-dimensional input and output data definition for the ANN}
\begin{threeparttable}
\begin{tabular*}{\textwidth}{c @{\extracolsep{\fill}} cccc}
\toprule
Variable type  &Raw definition & Normalize factor   &  Normalized form\tnote{$\star$}    &  Description  \\
\hline
\\[-8pt]
Input feature  &$k$            & $0.5u_iu_i$  \tnote{$\dagger$}  &  $50k/(50k+u_iu_i)$ \tnote{$\ddagger$}   &  turbulence intensity\\
Input feature  &$k/\omega$     & $\nu$                    &  $k/(k+50\nu\omega)$              &  estimated eddy viscosity\\
Label data     &$\nu_t$        & $k/\omega$               &  $5\nu_t/(5\nu_t+3k/\omega)$      &  true eddy viscosity\\
\bottomrule          
\end{tabular*}
\begin{tablenotes}
  \footnotesize
  \item[$\star$] The normalization strategy~\cite{liu2022on} takes a relative division $\bar{q_i}=|q_i|/(|{q_i|}+|\tilde{q_i}|)$, where $\bar{q_i}$ is the normalized variable, $q_i$ is the raw variable, and $\tilde{q_i}$ is the normalize factor. 
  \item[$\ddagger$] The value of $q_i$ and $\tilde{q_i}$ are adjusted with constant multiplier, ensuring $\bar{q_i}$ a uniform distribution between 0 and 1 within the range of training data.
  \item[$\dagger$] $0.5u_iu_i$ is the kinetic energy of mean flow, where $u_i$ is the velocity vector.
  \end{tablenotes}
\end{threeparttable}
\end{table}

\begin{table}[htbp!]
  \caption{ Hyper-parameters of the NN in ML-RANS framework \label{Table:RANSPara}}
  \centering
  \begin{tabular*}{\textwidth}
      {@{\hspace{6em}}c @{\extracolsep{\fill}}c @{\hspace{6em}}}
  \\[-11pt]
  \toprule 
  \textbf{Hyperparameters}     & \textbf{Recommend Values}  \\[1pt] \hline \\[-9pt]
  Number of Hidden Layers       & 3                         \\
  Number of Nodes per Layer     & 24                        \\
  Activation Function           & tanh                      \\
  Optimizer                     & Adam                      \\
  Learning Rate                 & 0.002                     \\
  L1 Regularization             & 4e-06                     \\
  L2 Regularization             & 6e-06                     \\
  Epoch Numbers                 & 5000                      \\
  \bottomrule  
  \end{tabular*}
\end{table}

\begin{figure}[!ht]
  \centering
  \includegraphics[width=1\textwidth,trim=40 17 43 60,clip]{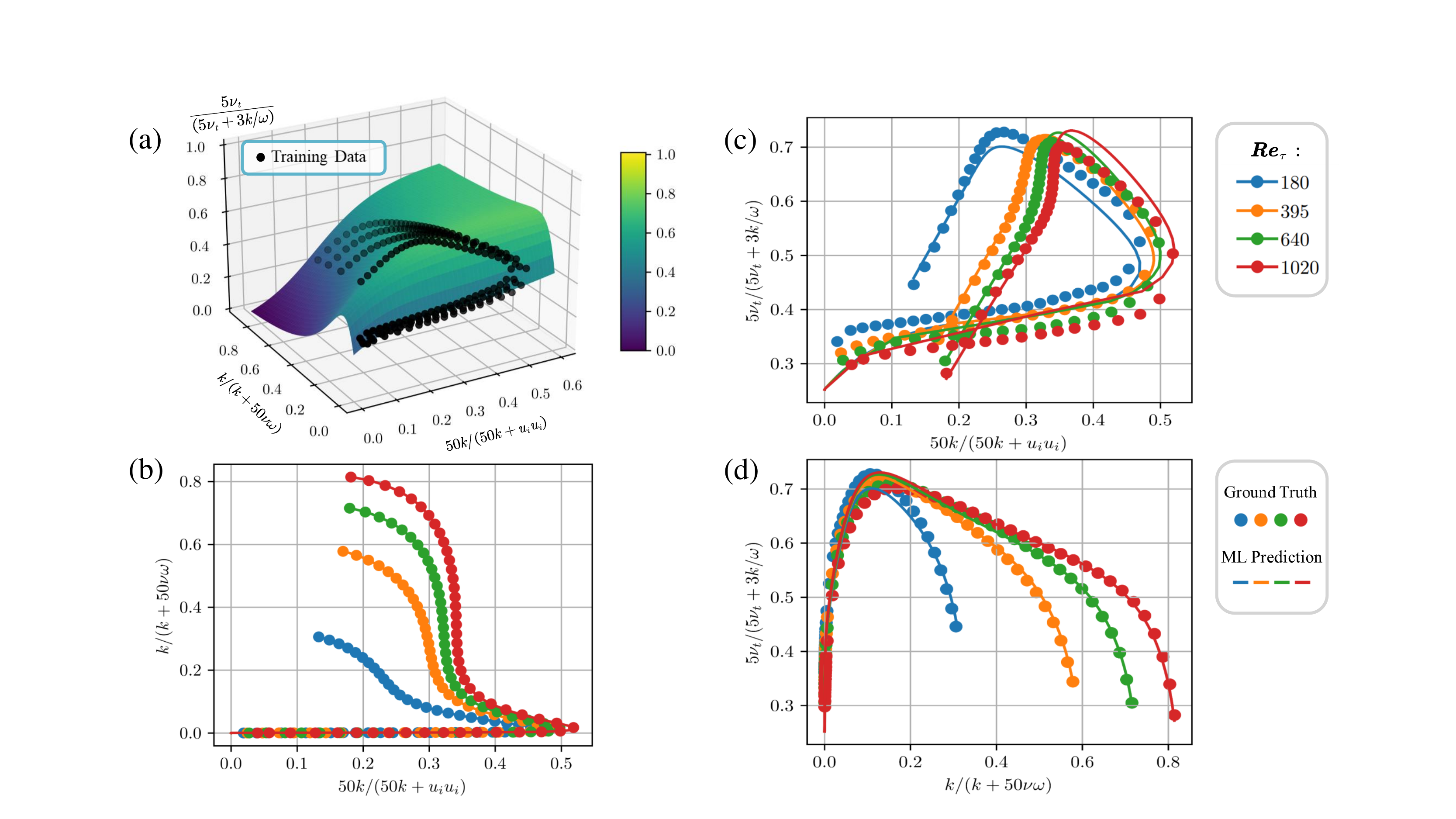}
  \caption{\label{fig:ML_RANS_Priori} \textit{A priori} performance of the ML model: (a) ML predicted surface and training data distribution in 3D frame: $x$-$y$ plane for input features, $z$ axis for output value. (b) 2D projections for $x$-$y$ plane of input features. (c) Comparison of ML predictions and training data in 2D projections for $x$-$z$ plane and (d) for $y$-$z$ plane. }
\end{figure}

The ML model's \textit{a priori} performance is shown in Fig.~\ref{fig:ML_RANS_Priori}. Fig.~\ref{fig:ML_RANS_Priori}.(a) shows the predicted curved surface of the NN compared with the training data, in which the $x$-axis stands for the first input feature, $\bar{q_1} = 50k/(50k+u_iu_i)$, and the $y$-axis is the second input feature, $\bar{q_2} = k/(k+50\nu\omega)$. The outputted variable, $\bar{\nu_t} = 5\nu_t/(5\nu_t+3k/\omega)$, is the $z$-axis. Fig.~\ref{fig:ML_RANS_Priori}.(b), (c), and (d) further show the 2D projection of (a) along the $z$, $y$, and $x$-axes, in which the output from the ML model is compared with the ground truth under the same input data. From the \textit{a priori} results, we can see that the NN finally obtains a surface with the training data falling on closely, and the smoothness and flatness of the surface are satisfying. The output's variation patterns with respect to each independent input feature are also accurately captured, as the predicted lines perfectly matched the training data in the $y$-$z$ plane (Fig.~\ref{fig:ML_RANS_Priori}.(d)). In the $x$-$z$ plane, the ML returns the averaged curve of the training data in the bottom half of the figure. This is because the inputs of the four cases overlap in this region (shown as an overlapped straight line in Fig.~\ref{fig:ML_RANS_Priori}(b) at $y \approx 0$), which does not provide enough separated distance to obtain distinguished outputs.

The trained model is then deployed into OpenFOAM and CFL3D through the C++ and Fortran APIs of the predictor library. The eddy-viscosity calculated from the $k$--$\omega$ SST model in the two CFD codes is then substituted with the ML model's prediction at each iteration step. The iteration stops until the residuals of the mean flow field, turbulence quantities, and the closure term reach certain criteria. The \textit{a posteriori} simulations are conducted at $Re_\tau=\left\{180, 395, 640, 830, 1020, 1340, 1670, 2000\right\}$, where cases at $Re_\tau=\left\{180, 395, 640, 1020\right\}$ and $Re_\tau=2000$, are compared with DNS data~\cite{RN62, abe2004surface} to evaluate the model's interpolation and extrapolation performance, respectively. The results at intervening Reynolds numbers (i.e., $Re_\tau=\left\{830, 1340, 1670,\right\}$) indicates that the ML model works properly within a certain range. 

\begin{figure}[!ht]
  \centering
  \includegraphics[width=0.91\textwidth,trim=50 0 50 8,clip]{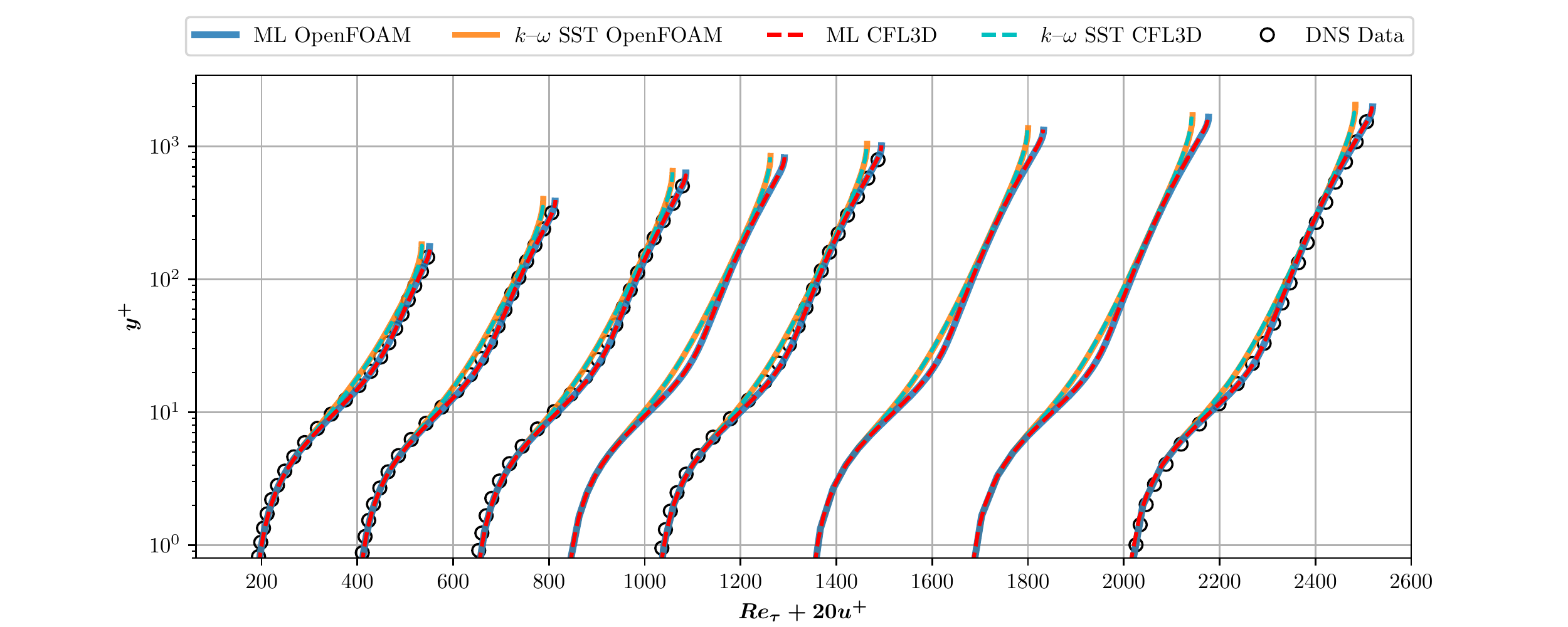}
  \vspace*{-10pt}
  \caption{\label{fig:ChanUPYPList} The velocity profiles of the series \textit{a posteriori} simulations.}
\end{figure}

\begin{figure}[!ht]
  \centering
  \includegraphics[width=0.91\textwidth,trim=50 0 50 32,clip]{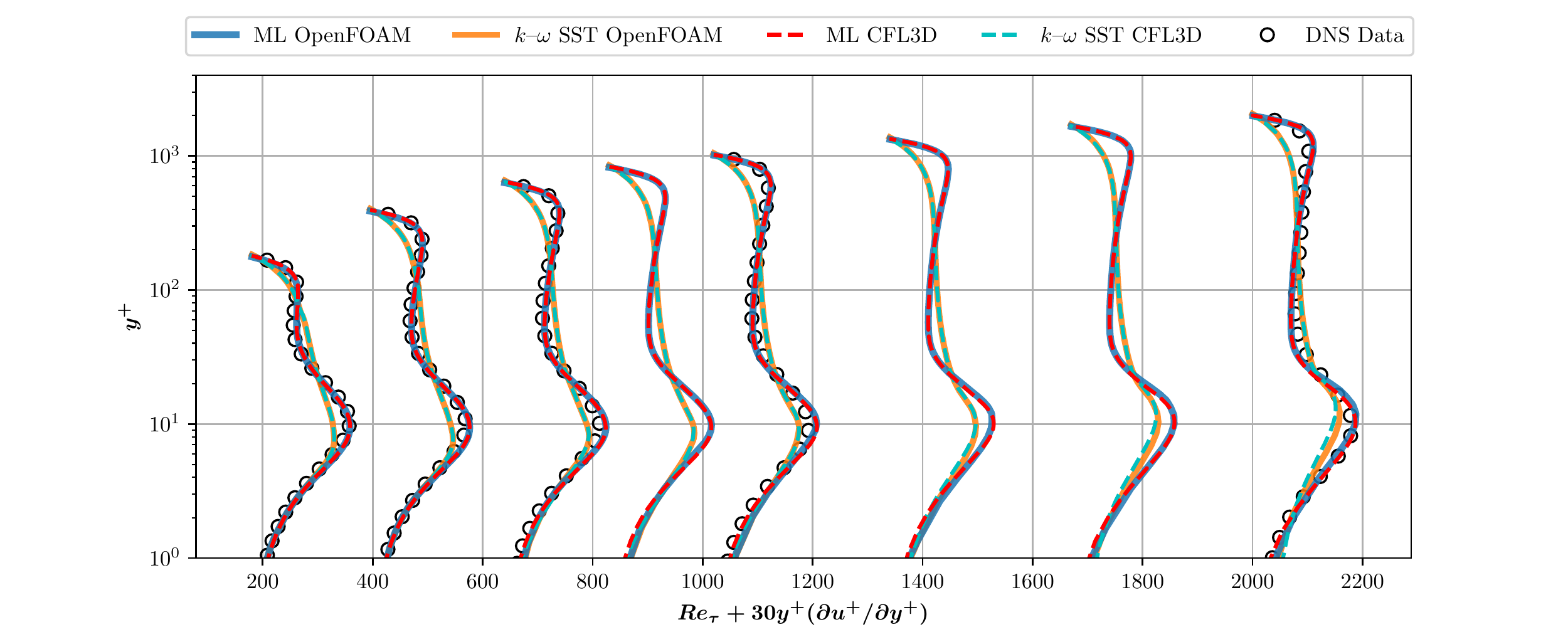}
  \vspace*{-10pt}
  \caption{\label{fig:ChanYDUDYList} The $y^+\partial u^+ / \partial y^+$ profiles of the series \textit{a posteriori} simulations.}
\end{figure}

\begin{figure}[!h]
  \centering
  \includegraphics[width=0.91\textwidth,trim=50 0 50 32,clip]{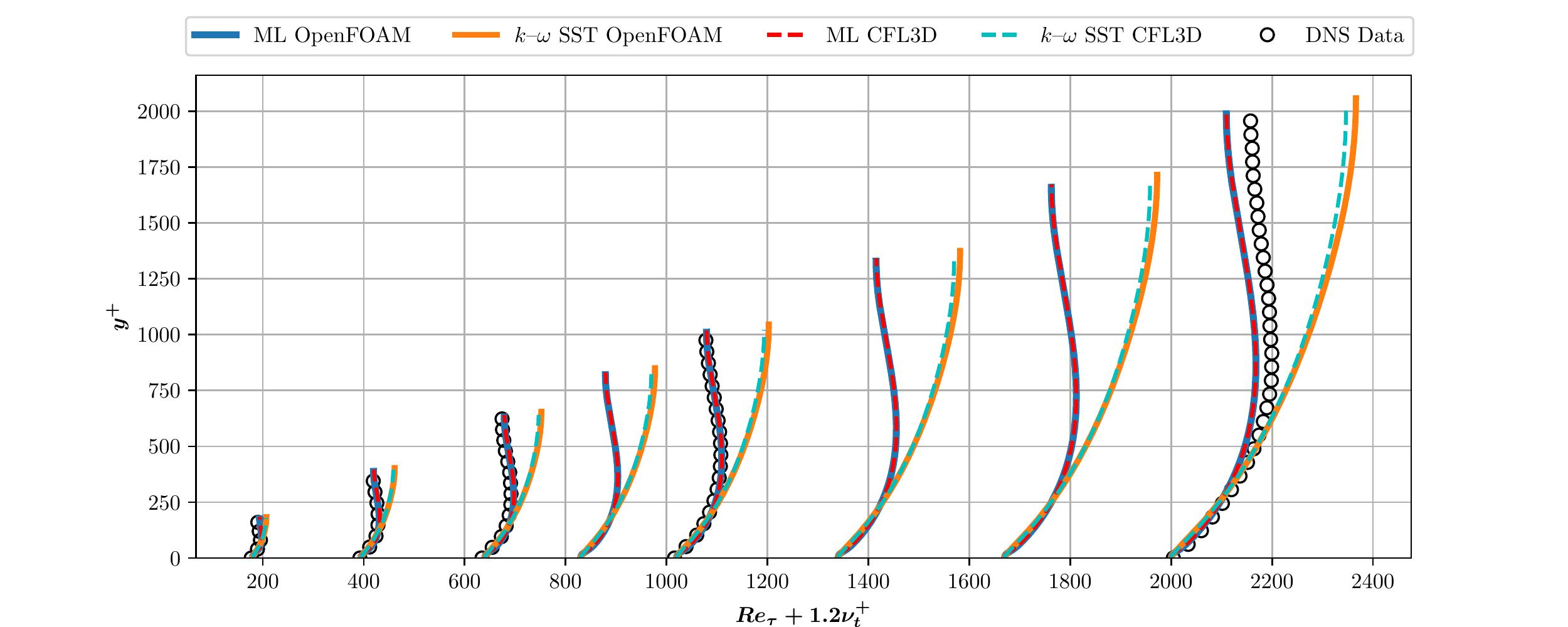}
  \vspace*{-10pt}
  \caption{\label{fig:ChanNUTList} The eddy-viscosity profiles of the series \textit{a posteriori} simulations.}
\end{figure}

\begin{figure}[!h]
  \centering
  \includegraphics[width=0.72\textwidth,trim=20 10 20 42,clip]{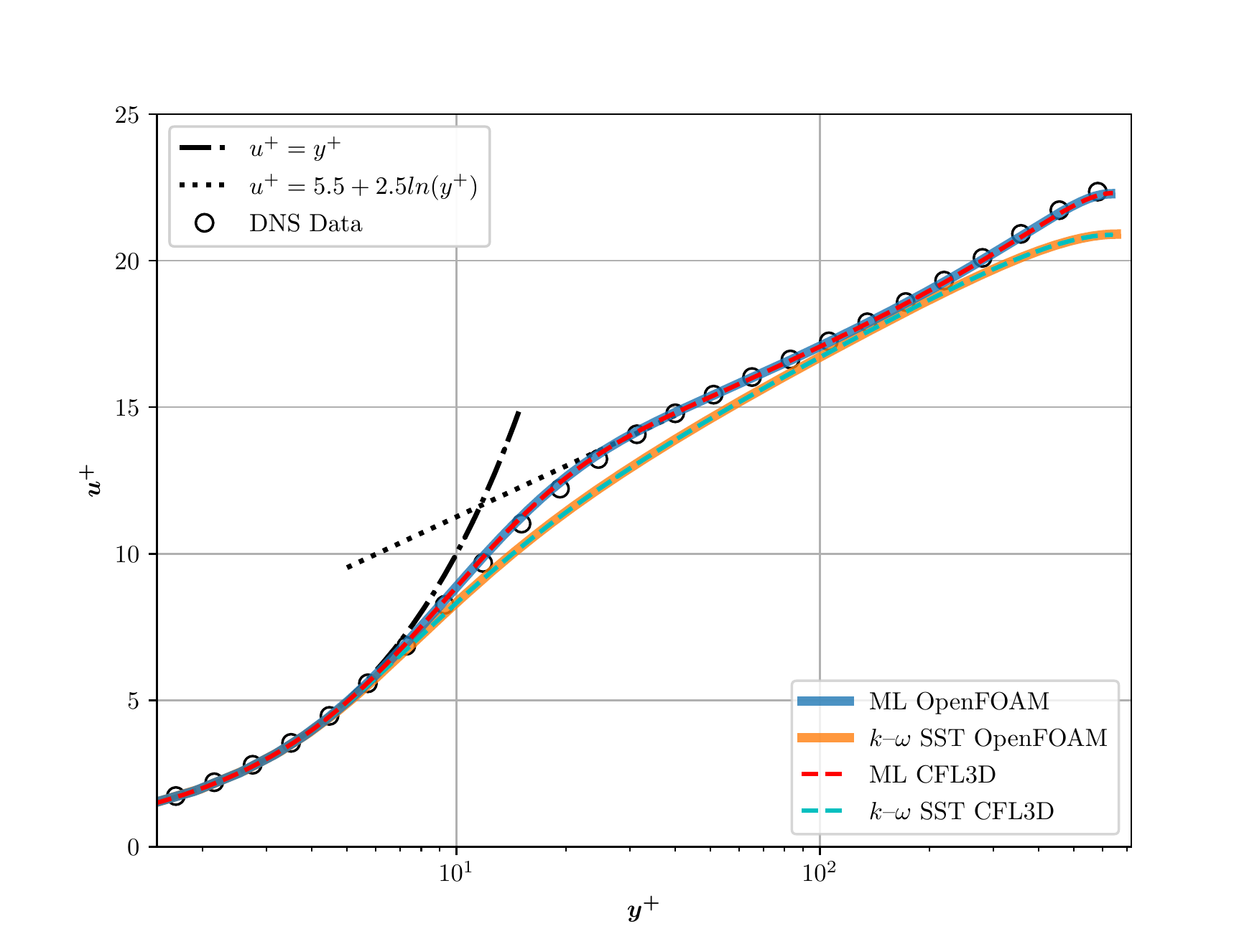}
  \vspace*{-10pt}
  \caption{\label{fig:Chan640UPYP} The velocity profile of the \textit{a posteriori} simulation at $Re_\tau=640$.}
\end{figure}

\begin{figure}[!h]
  \centering
  \includegraphics[width=1\textwidth,trim=50 0 50 27,clip]{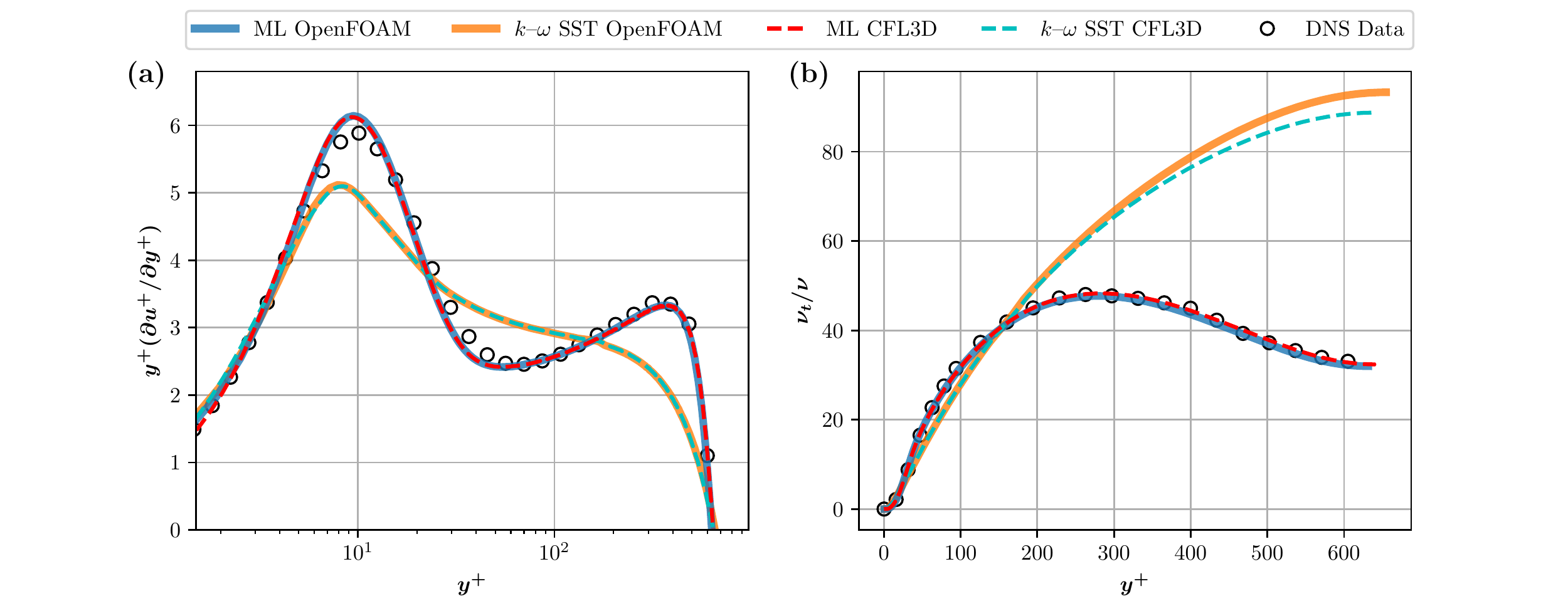}
  \vspace*{-15pt}
  \caption{\label{fig:Chan640ETC} The $y^+\partial u^+/\partial y^+$ and eddy-viscosity profiles of the \textit{a posteriori} simulation at $Re_\tau=640$.}
\end{figure}

This series of cases adopt a quasi-2D grid with 1153 $\times$ 101 nodes in the wall-normal and streamwise direction respectively, which ensures the grid resolution at the wall, $y_1^{+} \leq 1$, at the highest Reynolds number (i.e., $Re_\tau=2000$). The results from different models and CFD codes are compared in Figs.~\ref{fig:ChanUPYPList},~\ref{fig:ChanYDUDYList}, and~\ref{fig:ChanNUTList}, which respectively present profiles of velocity, velocity gradient, and eddy-viscosity. More specifically, the results from the case at $Re_\tau=640$ are shown separately to highlight the performance of the ML model in Figs.~\ref{fig:Chan640UPYP} and ~\ref{fig:Chan640ETC}. 

It can be seen from Figs.~\ref{fig:ChanUPYPList},~\ref{fig:ChanYDUDYList}, and~\ref{fig:ChanNUTList} that the results of the ML model from two different CFD codes overlap perfectly with each other, meaning that the implementation of the ML library is code-independent. The ML model presents improved results against the conventional $k$--$\omega$ SST model, which agrees with the previous result of Liu \textit{et al.}~\cite{liu2021an, liu2022on}. Although the implementation of the CFD algorithms differs significantly between the two CFD codes, the reproducibility of the ML-RANS framework is adequately demonstrated through the predictor library. 

\begin{figure}[!h]
  \centering
  \includegraphics[width=0.6\textwidth,trim=0 0 0 32,clip]{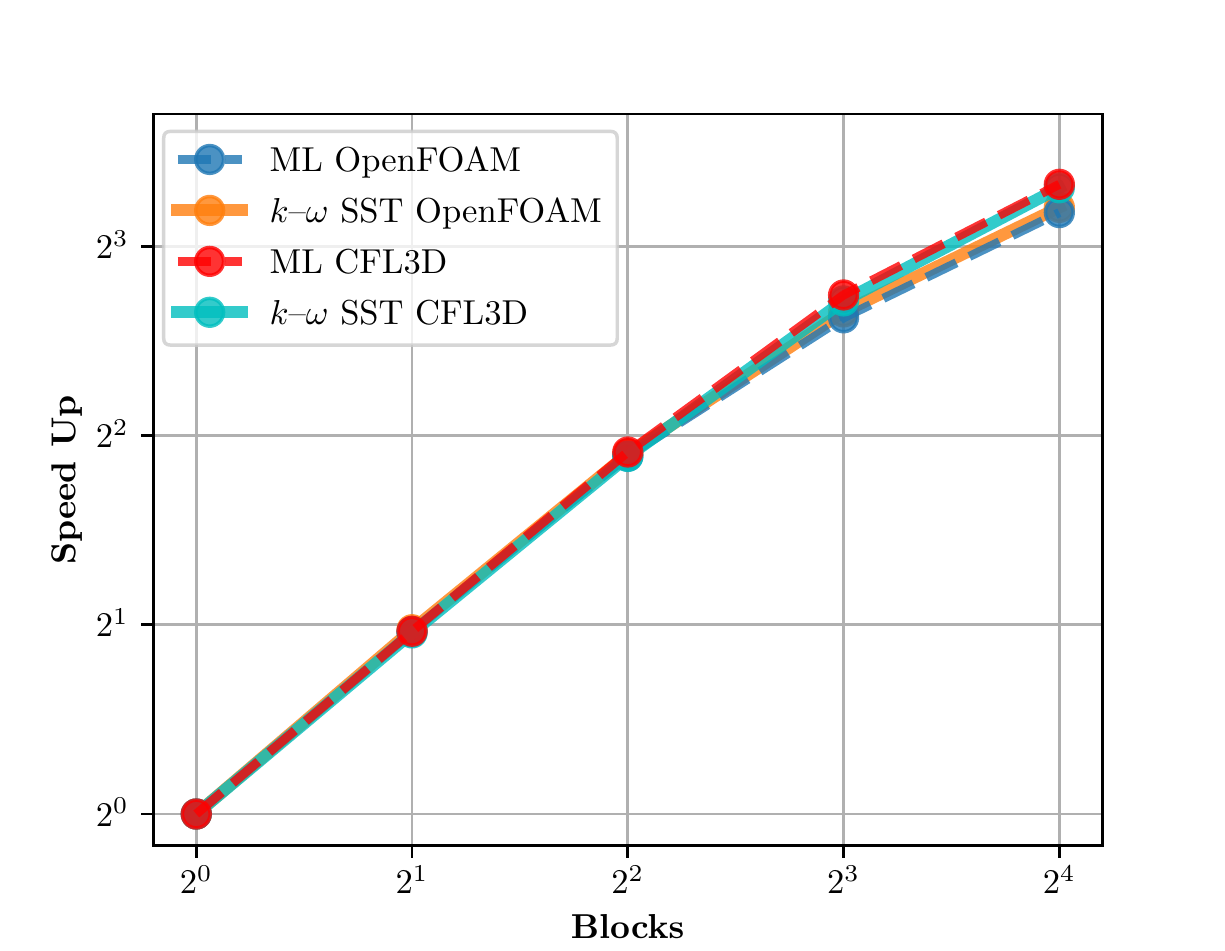}
  \vspace*{-8pt}
  \caption{\label{fig:ParallelSpeed} The parallel speed-up ratio of ML model and $k$--$\omega$ SST model on OpenFOAM and CFL3D.}
\end{figure}

The predictor library can be deployed to parallel CFD solvers based on domain decomposition and communication using the message passing interface (MPI). Under this parallel strategy, each thread will load an independent predictor to handle the ML prediction within its local subdomain. To evaluate the predictor's influence on the parallel performance, the simulation of channel flow with 1-16 MPI threads is conducted. The elapsed time in a single iteration and the extra time cost from the ML prediction in both CFD codes are shown in Table.~\ref{Table:ParallelTime}, from which we note that incorporation of the ML model brings an extra 10\% -- 15\% time cost compared with a conventional turbulence model. The speed-up ratio is demonstrated in Fig.~\ref{fig:ParallelSpeed}, where no obvious difference between the ML model and $k$--$\omega$ SST model is observed. Therefore, the parallel efficiency is not influenced by incorporating the predictor library.

\begin{table}[h!]
  \caption{Elapsed time in a single iteration for $k$--$\omega$ SST and ML model in channel flow under parallel configuration\label{Table:ParallelTime}}
  \centering
  \begin{threeparttable}
  \begin{tabular*}{\textwidth}{l @{\extracolsep{\fill}}cccccc}
  \toprule 
  \multirow{2}{*}{\textbf{Blocks}} & \multicolumn{3}{c}{\textbf{OpenFOAM}} & \multicolumn{3}{c}{\textbf{CFL3D \tnote{*}}} \\ \cline{2-4}  \cline{5-7} \\[-10pt]
                         & ML Model  & $k$--$\omega$ SST & ML Cost \tnote{**} & ML Model & $k$--$\omega$ SST & ML Cost \tnote{**}\\ \hline \\[-10pt]
  1                      & 0.308s    & 0.277s & 11.19\% & 0.422s   & 0.367s & 14.99\% \\
  2                      & 0.158s    & 0.141s & 11.90\% & 0.216s   & 0.189s & 14.55\% \\
  4                      & 0.083s    & 0.074s & 12.16\% & 0.112s   & 0.099s & 13.56\% \\
  8                      & 0.050s    & 0.044s & 13.64\% & 0.063s   & 0.056s & 12.76\% \\
  16                     & 0.034s    & 0.030s & 14.33\% & 0.042s   & 0.037s & 12.31\% \\ 
  \bottomrule
\end{tabular*}
\begin{tablenotes}
  \footnotesize
  \item[ *] CFL3D needs one extra core as the host of MPI slaves, for example, 17 cores are needed in the 16-block case.
  \item[**] The extra percentage of time cost compared with $k$--$\omega$ SST model.
  \end{tablenotes}
\end{threeparttable}
\end{table}

\section{Conclusion and Further Improvements\label{Sec:Conclusion}}

A predictor library is developed to deploy NN to CFD software conveniently. Using the Pimpl strategy, the library isolates the implementation details and achieves a pure-method class that can be easily extended to other languages. The library currently provides language-level APIs for C++ and Fortran, and high-level modules for OpenFOAM and CFL3D. The simplified APIs for loading models, setting the parallel options, and the data I/O functions enhanced the compatibility with most CFD programs. 

The basic usage of the library is first demonstrated through a simple \textbf{A + B} example, and further explained in modeling a 1-D heat transfer problem. The library is then applied in OpenFOAM and CFL3D to simulate turbulent channel flow using the ML model accounting for the effect of turbulence. The results are shown to be independent of CFD software. The incorporation of the library does not influence the parallel efficiency of a CFD code.
This newly developed ML predictor library could largely promote the deployment of ML models in mainstream CFD software.

Further improvement of the predictor can be made in three aspects. First, the library can be applied to other types of partial differential equation solvers for wider communities. Second, more backends should be supported, such as loading PyTorch and driving a Python interpreter, to provide more flexible ML support. Finally, the ML acceleration can be adapted to different parallel structures. For example, the ML prediction can be executed on GPU in a server-client mode in a heterogeneous computing system.

\section*{Acknowledgement}

We would like to acknowledge the UK
Engineering and Physical Sciences Research Council (EPSRC) through the Computational Science Centre for
Research Communities (CoSeC), and the UK Turbulence Consortium (no. EP/R029326/1).


\appendix
\section{Installation Guide\label{Append:Compile}}
\subsection*{C++ Core Predictor:}
\noindent
In the \href{https://github.com/Weishuo93/NN_Pred/tree/master/Predictor-Core/third_party}{\texttt{third\_party}} directory of the code repository, a bash script for downloading the necessary dependencies is available: \href{https://github.com/Weishuo93/NN_Pred/blob/master/Predictor-Core/third_party/DownloadThirdParty.sh}{\texttt{DownloadThirdParty.sh}}. Users can simply execute the script:
\begin{lstlisting}[language=bash,style=bashstyle]
$ cd Predictor-Core/third_party
$ sh DownloadThirdParty.sh
\end{lstlisting}
in which, the major dependencies are \textbf{TensorFlow C API} binary library and \textbf{Eigen} numerical template library. The former is the core functioning library to call ML prediction whereas the latter could enable efficient I/O option with SIMD (Single Instruction/Multiple Data)~\cite{amiri2020simd} operations. One can also download the dependencies with the following links (no need to compile the dependencies) manually:
\begin{itemize}
  \item \href{https://www.tensorflow.org/install/lang_c}{libtensorflow.so} 
  \item \href{https://gitlab.com/libeigen/eigen/-/releases}{Eigen} 
\end{itemize}
Two environment variables: \texttt{MY\_TF\_HOME} and \texttt{MY\_EIGEN\_HOME} are needed to specify the location of these third-party libraries. The bash script: \texttt{activate.sh} could quickly set the environment variables if the download script is executed:
\begin{lstlisting}[language=bash,style=bashstyle]
$ source activate.sh
\end{lstlisting}
Otherwise users needs to set them manually via:
\begin{lstlisting}[language=bash,style=bashstyle]
$ export MY_EIGEN_HOME=path/to/your/eigen/dir
$ export MY_TF_HOME=path/to/your/TF/dir/
\end{lstlisting}
Then the predictor library is ready to be compiled via (tested passed by GNU compiler ver 7.3.1):
\begin{lstlisting}[language=bash,style=bashstyle]
$ make cxxso      # C++ library
$ make cxxtest    # C++ test program loading A+B
$ make run        # run the test program
\end{lstlisting}
In addtion, two macro flags can be specified in the \texttt{Makefile} to achieve special usage:
\begin{lstlisting}[language=bash,style=bashstyle]
-D_CPU_SERIAL     # default run serially, otherwise will use all available CPUs
-D_DEBUG          # print error messages if not properly used
\end{lstlisting}
\subsection*{Fortran Interface:}
\noindent
With C++ core predictor compiled, one with GNU Fortran compiler~\cite{reid2007new} could directly compile the Fortran interface and its test program via:
\begin{lstlisting}[language=bash,style=bashstyle]
$ make f90so      # Fortran predictor library
$ make f90test    # Fortran test program loading A+B
$ make runf       # run the Fortran test program
\end{lstlisting}
\subsection*{OpenFOAM Module:}
\noindent
Apart from the C++ core predictor library, the OpenFOAM module needs to be compiled within an activated \texttt{wmake} environment, which means OpenFOAM installation is required first. The module supports OpenFOAM version $\geq$ 4, and can be compiled and tested via:
\begin{lstlisting}[language=bash,style=bashstyle]
$ cd OpenFOAM-Extension  # Go to the OF Extension dir
$ make ofso       # OpenFOAM module library
$ make oftest     # OpenFOAM test program loading A+B
$ make runof      # run the OpenFOAM test program
\end{lstlisting}

\subsection*{Fortran CFD Module (i.e., the CFL3D module):}
\noindent
The self-developed CFL3D solver with ML taking over the prediction of eddy-viscosity can be compiled via:
\begin{lstlisting}[language=bash,style=bashstyle]
$ cd CFL3D-Extension  # Go to the OF Extension dir
$ make cfl3d_seq_ori    # Original solver serial version
$ make cfl3d_mpi_ori    # Original solver parallel version
$ make cfl3d_seq_ml     # ML-RANS solver serial version
$ make cfl3d_mpi_ml     # ML-RANS solver parallel version
\end{lstlisting}
Instead of compiling the module into an independent library, \href{https://github.com/Weishuo93/NN_Pred/blob/master/CFL3D-Extension/Solvers/Code_Template/ML_Template_Module.F90}{a Fortran 90 format module template} is provided to compile with CFD program together. Users are recommended to modify the template following the instructions. This template module design pattern is tested as compatible with fixed format (\texttt{*.F}) and free format (\texttt{*.F90}) because it only interacts with the arrays without allocating extra memory spaces in the main CFD program. All the other intermediate arrays and class instances are maintained by the module itself. Therefore the risk of memory leak in a legacy Fortran program can be avoided. 

\section{Detailed API Documents\label{Append:APIs}}
This appendix records the entire document of the APIs in C++, Fortran, and OpenFOAM extension. The extension module for CFL3D is shown as a template to provide a solution dealing with legacy CFD code. The same \textbf{A + B} model is adopted in examples for C++, Fortran, and OpenFOAM for a better understanding. For C++ and Fortran, the alternative APIs are listed after the code example with the specified step ID for a more customized usage. 

The two different formats of TensorFlow model can be generated by simply executing \href{https://github.com/Weishuo93/NN_Pred/blob/master/Predictor-Core/test/models/createmodel_AplusB.py}{\texttt{createmodel\_AplusB.py}}, where the TensorFlow version is automatically detected to match the API difference between TensorFlow version 1 and 2:
\begin{lstlisting}[language=bash,style=bashstyle]
$ python3 createmodel_AplusB.py PB          # Save model in Protobuf format 
$ python3 createmodel_AplusB.py SavedModel  # Save model in SavedModel format 
\end{lstlisting}
The model will finally be saved with the following filename: 
\begin{lstlisting}[language=bash,style=bashstyle]
simple_graph_tf1.pb               # TF1 saved *.pb format 
simple_graph_tf2.pb               # TF2 saved *.pb format 
saved_tf1_model                   # TF1 SavedModel format (Normally by Estimator)
saved_tf2_model                   # TF2 SavedModel format (Normally by TF2 Keras)
\end{lstlisting}

\subsection{C++ Core Predictor Class APIs}

The C++ code example deploying the \textbf{A + B} model is shown in Listing~\ref{lst:AplusB_Append}, in which, the necessary steps, with their function prototypes and locations in the code example, are marked in Table.~\ref{Table:CppLoc}: 

\begin{table}[h!]
  \small
  \caption{ID, function prototypes and locations of necessary steps in C++\label{Table:CppLoc}}
  \centering
  \begin{threeparttable}
    \begin{tabular*}{\textwidth}{cllc}
      \toprule 
      \textbf{Step ID} & \textbf{Usage}  & \textbf{Function Prototype}   \tnote{*}                   & \textbf{Location} \\ \hline
      C01         & Load Model      & \verb|Predictor(std::string PBfile)|                                     & 06-07    \\
      C02         & Register Nodes  & \verb|regist_node(std::string node_name, Predictor::NodeType tp)|        & 09-16    \\
      C03         & Set Data Counts & \verb|set_data_count(int n_data)|                                        & 18-19    \\
      C04         & Set Input Data  & \verb|set_node_data(std::string node_name, std::vector<T>& data)|        & 28-30    \\
      C05         & Run Model       & \verb|run()|                                                             & 32-33    \\
      C06         & Get Output Data & \verb|get_node_data(std::string node_name, std::vector<T>& data)|        & 35-36    \\
      \bottomrule  
    \end{tabular*}
    \begin{tablenotes}
     \footnotesize
     \item[*] The function prototypes are all under public class: \verb|class Predictor|. Therefore, the full reference of functions, for example for C01, should be: \verb|Predictor::Predictor(std::string PBfile)|
     \end{tablenotes}
  \end{threeparttable}
\end{table}
\noindent
Basically, the code example is similar to the code shown in Sec.~\ref{Sec:CppPredictor}, the only difference is that the output array is not transposed in this case, which reads:
\begin{lstlisting}[language=C++,style=mystyle,caption=Minimal example to run A + B model in C++, label={lst:AplusB_Append}]
// Header files
#include "predictor.h"   // Predictor header
#include <vector>        // C++ standard header

int main(int argc, char const *argv[]) {
    // Load Model:
    Predictor pd("simple_graph_tf2.pb"); // Model's path or filename

    // Register node:
    // Inputs: 
    // Predictor::INPUT_NODE is the node type enumerate 
    pd.regist_node("input_a", Predictor::INPUT_NODE);
    pd.regist_node("input_b", Predictor::INPUT_NODE);
    // Outputs:
    // Predictor::OUTPUT_NODE is the node type enumerate 
    pd.regist_node("result", Predictor::OUTPUT_NODE);

    // Set the number of data instances (n=3)
    pd.set_data_count(3);

    // Create external source of input/output data array:
    // Inputs:
    std::vector<float> vec_input1_float = {1.1, 2.2, 3.3, 4.4, 5.5, 6.6};
    std::vector<int> vec_input2_float = {6, 5, 4, 3, 2, 1};
    // Outputs:
    std::vector<double> vec_out_float(6);

    // Set data for input nodes
    pd.set_node_data("input_a", vec_input1);
    pd.set_node_data("input_b", vec_input2);

    // Run model
    pd.run();

    // Get output into the target container
    pd.get_node_data("result", vec_out);
    
    // Check results, expected calculation results:
    // vec_out: [7.1, 7.3, 7.5, 7.2, 7.4, 7.6]
    //          [C11, C21, C31, C12, C22, C32]    
    return 0;
}
\end{lstlisting}
\noindent
\textbf{Entire API list for C++}:
\small
\begin{itemize}
  \item \textbf{C01: Loading Model:} 
  \begin{itemize}
    \item \verb|Predictor(std::string pbfile)|
    \begin{itemize}
      \item \verb|pbfile| -- the file name of the PB graph (i.e.,\verb|simple_graph_tf2.pb|).
    \end{itemize}
    
    Class constructor, to create the \verb|Predictor| object from a \verb|*.pb| format.
    \item \verb|Predictor(std::string folder, std::string tag)|
    \begin{itemize}
      \item \verb|folder| -- the directory of the \verb|SavedModel| format (this format itself is a folder).
      \item \verb|tags|   -- tags label within a \verb|SavedModel| format, by default is \verb|serve|\footnote[6]{Users could use the \href{https://www.tensorflow.org/guide/saved_model}{\texttt{saved\_model\_cli}} to find out the tags in a \texttt{SavedModel} format}.
    \end{itemize}
    
    Class constructor, to create the \verb|Predictor| object from a \verb|SavedModel| format. 
    \item \verb|Predictor(std::string pbfile, uint8_t para_intra, uint8_t para_inter)|
    \begin{itemize}
      \item \verb|pbfile| -- the name of the PB graph (i.e.,\verb|simple_graph_tf2.pb|).
      \item \verb|para_intra| -- number of threads for internal parallelization (e.g., matrix multiplication and reduce sum).
      \item \verb|para_inter| -- number of threads for operations independent with each other.
    \end{itemize}
    
    Class constructor, to create the \verb|Predictor| object from a \verb|*.pb| format with parallelization configs. If both \verb|para_intra| and \verb|para_inter| are set to be \textbf{0}, the system will pick an appropriate number. If they are set to be \textbf{1}, the predictor will run model serially (an easy way to cooperate with MPI pattern in CFD codes). 
    \item \verb|Predictor(std::string folder, std::string tag, uint8_t para_intra, uint8_t para_inter)|
    \begin{itemize}
      \item \verb|folder| -- the directory of the \verb|SavedModel| format (this format itself is a folder).
      \item \verb|tags|   -- tags label within a \verb|SavedModel| format, by default is \verb|serve|.
      \item \verb|para_intra| -- number of threads for internal parallelization (e.g., matrix multiplication and reduce sum).
      \item \verb|para_inter| -- number of threads for operations independent with each other.
    \end{itemize}
    
    Class constructor, to create the \verb|Predictor| object from a \verb|SavedModel| format with parallelization configs. If both \verb|para_intra| and \verb|para_inter| are set to be \textbf{0}, the system will pick an appropriate number. If they are set to be \textbf{1}, the predictor will run model serially (an easy way to cooperate with MPI pattern in CFD codes). 
  \end{itemize}

  \item \textbf{C02: Register Nodes:} 
  \begin{itemize}
    \item \verb|regist_node(std::string node_name, NodeType type)|
    \begin{itemize}
    \item \verb|node_name| -- the name of the node to be registered as input or output.
    \item \verb|type| -- the enumerate to specify node type, the available options are:
      \begin{itemize}
      \item \verb|Predictor::INPUT_NODE| -- to register node as input.
      \item \verb|Predictor::OUTPUT_NODE| -- to register node as output.
      \end{itemize}
    \end{itemize}
    
    To register input and output node for feeding and extracting data.
  \end{itemize}

  \item \textbf{C03: Set Data Counts:} 
  \begin{itemize}
    \item \verb|set_data_count(int n_data)|
    \begin{itemize}
    \item \verb|n_data| -- the number of data instances.
    \end{itemize}
    
    To set the number of data instances to substitute the unknown dimension of input/output tensors. For example, in the A + B example, the input/output shapes are all [-1, 2], the function could set the unknown -1 into concrete value so that the inner data containers can be created.
  \end{itemize}

  \item \textbf{C04: Set Input Data:} 
  \begin{itemize}
    \item \verb|set_node_data(std::string node_name, std::vector<T>& data)|
    \begin{itemize} 
    \item \verb|node_name| -- the name of the input node to be fed with external data.
    \item \verb|data| -- the external data defined with C++ STL library (i.e., \verb|std::vector|)
    \end{itemize}
    
    To feed the internal input data container registered under \verb|node_name| with external data source hold by a C++ standard template library (STL) \verb|std::vector|. If the data type of the internal container and external source are different, this function would automatically cast the datatype to resolve the difference.

    \item \verb|set_node_data(std::string node_name, std::vector<T>& data, DataLayout layout)|
    \begin{itemize} 
    \item \verb|node_name| -- the name of the input node to be fed with external data.
    \item \verb|data| -- the external data defined with C++ STL library (i.e., \verb|std::vector|)
    \item \verb|layout| -- the enumerate to specify whether the memory layout is transposed, the available options are:         
      \begin{itemize}
        \item \verb|Predictor::RowMajor| -- to hold the original memory sequence of the external data source.
        \item \verb|Predictor::ColumnMajor| -- to perform matrix transpose while set data to internal container.
      \end{itemize}
    \end{itemize}
    
    To feed the internal input data container registered under \verb|node_name| with external data source hold by a C++ standard template library (STL) \verb|std::vector|. If the data type of the internal container and external source are different, this function would automatically cast the datatype to resolve the difference. The memory layout would be changed during the mapping process if \verb|Predictor::ColumnMajor| is passed to the function. 

    \item \verb|set_node_data(std::string node_name, std::vector<T>& data, DataLayout layout, |\\           
    \verb|CopyMethod method)|
    \begin{itemize} 
    \item \verb|node_name| -- the name of the input node to be fed with external data.
    \item \verb|data| -- the external data defined with C++ STL library (i.e., \verb|std::vector|)
    \item \verb|layout| -- the enumerate to specify whether the memory layout is transposed, the available options are:         
      \begin{itemize}
        \item \verb|Predictor::RowMajor| -- to hold the original memory sequence of the external data source.
        \item \verb|Predictor::ColumnMajor| -- to perform matrix transpose while set data to internal container.
      \end{itemize}
    \item \verb|method| -- the enumerate to specify the copy method to the internal container:         
      \begin{itemize}
        \item \verb|Predictor::Simple| -- to copy/cast the arrays element-wise through C++ loop (possibly cache miss).
        \item \verb|Predictor::Eigen| -- to copy/cast the arrays via \verb|Eigen| library (SIMD is enabled).
      \end{itemize}
    \end{itemize}
    
    To feed the internal input data container registered under \verb|node_name| with external data source hold by a C++ standard template library (STL) \verb|std::vector|. If the data types of the internal container and external source are different, this function would automatically cast the datatype to resolve the difference. The memory layout would be changed during the mapping process if \verb|Predictor::ColumnMajor| is passed to the function. The copy/cast procedure could be accelerated via \verb|Eigen| library by specifying \verb|Predictor::Eigen| options.

    \item \verb|set_node_data(std::string node_name, T* p_data, int array_size)|
    \begin{itemize} 
    \item \verb|node_name| -- the name of the input node to be fed with external data.
    \item \verb|p_data| -- the pointer to the first element of external data array.
    \item \verb|array_size| -- the number of data elements of the external data array.
    \end{itemize}
    
    To feed the internal input data container registered under \verb|node_name| with external data source specified by the pointer to the array and the number of data elements. If the data types of the internal container and external source are different, this function would automatically cast the datatype to resolve the difference.

    \item \verb|set_node_data(std::string node_name, T* p_data, int array_size, DataLayout layout)|
    \begin{itemize} 
    \item \verb|node_name| -- the name of the input node to be fed with external data.
    \item \verb|p_data| -- the pointer to the first element of external data array.
    \item \verb|array_size| -- the number of data elements of the external data array.
    \item \verb|layout| -- the enumerate to specify whether the memory layout is transposed, the available options are:         
      \begin{itemize}
        \item \verb|Predictor::RowMajor| -- to hold the original memory sequence of the external data source.
        \item \verb|Predictor::ColumnMajor| -- to perform matrix transpose while set data to internal container.
      \end{itemize}
    \end{itemize}
    
    To feed the internal input data container registered under \verb|node_name| with external data source specified by the pointer to the array and the number of data elements. If the data types of the internal container and external source are different, this function would automatically cast the datatype to resolve the difference. The memory layout would be changed during the mapping process if \verb|Predictor::ColumnMajor| is passed to the function. 

    \item \verb|set_node_data(std::string node_name, T* p_data, int array_size, DataLayout layout, |\\           
    \verb|CopyMethod method)|
    \begin{itemize} 
    \item \verb|node_name| -- the name of the input node to be fed with external data.
    \item \verb|p_data| -- the pointer to the first element of external data array.
    \item \verb|array_size| -- the number of data elements of the external data array.
    \item \verb|layout| -- the enumerate to specify whether the memory layout is transposed, the available options are:         
      \begin{itemize}
        \item \verb|Predictor::RowMajor| -- to hold the original memory sequence of the external data source.
        \item \verb|Predictor::ColumnMajor| -- to perform matrix transpose while set data to internal container.
      \end{itemize}
    \item \verb|method| -- the enumerate to specify the copy method to the internal container:         
      \begin{itemize}
        \item \verb|Predictor::Simple| -- to copy/cast the arrays element-wise through C++ loop (possibly cache miss).
        \item \verb|Predictor::Eigen| -- to copy/cast the arrays via \verb|Eigen| library (SIMD is enabled).
      \end{itemize}
    \end{itemize}
    
    To feed the internal input data container registered under \verb|node_name| with external data source specified by the pointer to the array and the number of data elements. If the data types of the internal container and external source are different, this function would automatically cast the datatype to resolve the difference. The memory layout would be changed during the mapping process if \verb|Predictor::ColumnMajor| is passed to the function. The copy/cast procedure could be accelerated via \verb|Eigen| library by specifying \verb|Predictor::Eigen| options.
  \end{itemize}

  \item \textbf{C05: Run Model:} 
  \begin{itemize}
    \item \verb|run()|

    To run the model prediction, the result will be stored in the internal data container holding the model's output.
  \end{itemize}

  \item \textbf{C06: Get Output Data:} 
  \begin{itemize}
    \item \verb|get_node_data(std::string node_name, std::vector<T>& data)|
    \begin{itemize} 
    \item \verb|node_name| -- the name of the output node's data to be extracted to external array.
    \item \verb|data| -- the external data defined with C++ STL library (i.e., \verb|std::vector|)
    \end{itemize}
    
    To extract the data stored in the the internal output data container registered under \verb|node_name| into external data array hold by a C++ standard template library (STL) \verb|std::vector|. If the data types of the internal container and external source are different, this function would automatically cast the datatype to resolve the difference.

    \item \verb|get_node_data(std::string node_name, std::vector<T>& data, DataLayout layout)|
    \begin{itemize} 
    \item \verb|node_name| -- the name of the output node's data to be extracted to external array.
    \item \verb|data| -- the external data defined with C++ STL library (i.e., \verb|std::vector|)
    \item \verb|layout| -- the enumerate to specify whether the memory layout is transposed, the available options are:         
      \begin{itemize}
        \item \verb|Predictor::RowMajor| -- to hold the original memory sequence of the external data source.
        \item \verb|Predictor::ColumnMajor| -- to perform matrix transpose while set data to internal container.
      \end{itemize}
    \end{itemize}
    
    To extract the data stored in the the internal output data container registered under \verb|node_name| into external data array hold by a C++ standard template library (STL) \verb|std::vector|. If the data types of the internal container and external source are different, this function would automatically cast the datatype to resolve the difference. The memory layout would be changed during the mapping process if \verb|Predictor::ColumnMajor| is passed to the function. 

    \item \verb|get_node_data(std::string node_name, std::vector<T>& data, DataLayout layout, |\\           
    \verb|CopyMethod method)|
    \begin{itemize} 
    \item \verb|node_name| -- the name of the output node's data to be extracted to external array.
    \item \verb|data| -- the external data defined with C++ STL library (i.e., \verb|std::vector|)
    \item \verb|layout| -- the enumerate to specify whether the memory layout is transposed, the available options are:         
      \begin{itemize}
        \item \verb|Predictor::RowMajor| -- to hold the original memory sequence of the external data source.
        \item \verb|Predictor::ColumnMajor| -- to perform matrix transpose while set data to internal container.
      \end{itemize}
    \item \verb|method| -- the enumerate to specify the copy method to the internal container:         
      \begin{itemize}
        \item \verb|Predictor::Simple| -- to copy/cast the arrays element-wise through C++ loop (possibly cache miss).
        \item \verb|Predictor::Eigen| -- to copy/cast the arrays via \verb|Eigen| library (SIMD is enabled).
      \end{itemize}
    \end{itemize}
    
    To extract the data stored in the the internal output data container registered under \verb|node_name| into external data array hold by a C++ standard template library (STL) \verb|std::vector|. If the data type of the internal container and external source are different, this function would automatically cast the datatype to resolve the difference. The memory layout would be changed during the mapping process if \verb|Predictor::ColumnMajor| is passed to the function. The copy/cast procedure could be accelerated via \verb|Eigen| library by specifying \verb|Predictor::Eigen| options.

    \item \verb|get_node_data(std::string node_name, T* p_data, int array_size)|
    \begin{itemize} 
    \item \verb|node_name| -- the name of the output node's data to be extracted to external array.
    \item \verb|p_data| -- the pointer to the first element of external data array.
    \item \verb|array_size| -- the number of data elements of the external data array.
    \end{itemize}
    
    To extract the data stored in the the internal output data container registered under \verb|node_name| into external data array specified by the pointer to the array and the number of data elements. If the data type of the internal container and external source are different, this function would automatically cast the datatype to resolve the difference.

    \item \verb|get_node_data(std::string node_name, T* p_data, int array_size, DataLayout layout)|
    \begin{itemize} 
    \item \verb|node_name| -- the name of the output node's data to be extracted to external array.
    \item \verb|p_data| -- the pointer to the first element of external data array.
    \item \verb|array_size| -- the number of data elements of the external data array.
    \item \verb|layout| -- the enumerate to specify whether the memory layout is transposed, the available options are:         
      \begin{itemize}
        \item \verb|Predictor::RowMajor| -- to hold the original memory sequence of the external data source.
        \item \verb|Predictor::ColumnMajor| -- to perform matrix transpose while set data to internal container.
      \end{itemize}
    \end{itemize}
    
    To extract the data stored in the the internal output data container registered under \verb|node_name| into external data array specified by the pointer to the array and the number of data elements. If the data type of the internal container and external source are different, this function would automatically cast the datatype to resolve the difference. The memory layout would be changed during the mapping process if \verb|Predictor::ColumnMajor| is passed to the function. 

    \item \verb|get_node_data(std::string node_name, T* p_data, int array_size, DataLayout layout, |\\           
          \verb|              CopyMethod method)|
    \begin{itemize} 
    \item \verb|node_name| -- the name of the output node's data to be extracted to external array.
    \item \verb|p_data| -- the pointer to the first element of external data array.
    \item \verb|array_size| -- the number of data elements of the external data array.
    \item \verb|layout| -- the enumerate to specify whether the memory layout is transposed, the available options are:         
      \begin{itemize}
        \item \verb|Predictor::RowMajor| -- to hold the original memory sequence of the external data source.
        \item \verb|Predictor::ColumnMajor| -- to perform matrix transpose while set data to internal container.
      \end{itemize}
    \item \verb|method| -- the enumerate to specify the copy method to the internal container:         
      \begin{itemize}
        \item \verb|Predictor::Simple| -- to copy/cast the arrays element-wise through C++ loop (possibly cache miss).
        \item \verb|Predictor::Eigen| -- to copy/cast the arrays via \verb|Eigen| library (SIMD is enabled).
      \end{itemize}
    \end{itemize}
    
    To extract the data stored in the the internal output data container registered under \verb|node_name| into external data array specified by the pointer to the array and the number of data elements. If the data type of the internal container and external source are different, this function would automatically cast the datatype to resolve the difference. The memory layout would be changed during the mapping process if \verb|Predictor::ColumnMajor| is passed to the function. The copy/cast procedure could be accelerated via \verb|Eigen| library by specifying \verb|Predictor::Eigen| options.
  \end{itemize}

  \item \textbf{Auxiliary Functions:} 
  \begin{itemize}
    \item \verb|print_operations()|

    To print all the node information (type and shape) in the loaded model.
    \item \verb|print_operations(std::string node_name)|
    \begin{itemize} 
      \item \verb|node_name| -- the name of node to print shape and type information.
    \end{itemize} 

    To print the node information (type and shape) with specified name in the loaded model.
  \end{itemize}

\end{itemize}

\subsection{Fortran Calling the Predictor Function \label{Sec:FortranPredictor}}
The Fortran APIs is a wrapper of the C++ predictor class through \texttt{ISO\_C\_BINDING}, which is supposed to be a necessary component in a GNU Fortran compiler. A minimal code example that closely replicates the C++ calling sequence is shown below (written in \texttt{F90}). The major subroutines with their locations in the example are listed in Table.~\ref{Table:FortranLoc}. One difference between the C++ example is that in the data I/O process, the Fortran APIs adopt the pointer-and-data-number calling methods, since the Fortran's array shape information is not naturally passed to C++ via \texttt{ISO\_C\_BINDING}.

\begin{table}[h!]
  \small
  \caption{ID, function prototypes and locations of necessary steps in Fortran\label{Table:FortranLoc}}
  \centering
  \begin{threeparttable}
    \begin{tabular*}{\textwidth}{cllc}
      \toprule 
      \textbf{Step ID} & \textbf{Usage}  & \textbf{Function Prototype}                   & \textbf{Location} \\ \hline
      F01         & Load Model       & \verb|ptr = C_CreatePredictor(file_name)|   \tnote{*}                     & 13-14    \\
      F02         & Register Inputs  & \verb|C_PredictorRegisterInputNode(ptr, in_node_name)|                    & 16-18    \\
      F03         & Register Outputs & \verb|C_PredictorRegisterOutputNode(ptr, in_node_name)|                   & 20-21    \\
      F04         & Set Data Counts  & \verb|C_PredictorSetDataCount(ptr, n_data)|                               & 23-24    \\
      F05         & Set Input Data   & \verb|C_PredictorSetNodeData(ptr, in_node, input_arr , n_element)|        & 26-28    \\
      F06         & Run Model        & \verb|C_PredictorRun(ptr)|                                                & 30-31    \\
      F07         & Get Output Data  & \verb|C_PredictorGetNodeData(ptr, out_node, output_arr , n_element)|      & 33-34    \\
      F08         & Finalize Model   & \verb|C_DeletePredictor(ptr)|                                             & 39-40    \\
      \bottomrule  
    \end{tabular*}
    \begin{tablenotes}
     \footnotesize
     \item[*] The type of the returned value should be declared as: \verb|type(c_ptr) :: ptr|. 
     \end{tablenotes}
  \end{threeparttable}
\end{table}
\noindent
The Fortran code example to load the \textbf{A + B} model reads:
\begin{lstlisting}[language=Fortran,style=mystyle,caption=Minimal example to run A + B model in Fortran, label={lst:FortranAplusB}]
program main
use ml_predictor  !The predictor module that declared all the function interfaces
use iso_c_binding, only: c_ptr  !C-pointer to the Predictor instance

    implicit none
    type(c_ptr) :: ptr    ! C-pointer to the Predictor instance

    ! External input/output arrays from fortran program (Support up to 6d)
    real(kind=4), dimension(2,3)  :: arr_a = reshape((/0.0, 1.1, 2.2, 3.3, 4.4, 5.5/), (/2,3/))
    integer(kind=4), dimension(6) :: arr_b = (/5, 4, 3, 2, 1, 0/)
    real(kind=8), dimension(3,2)  :: arr_c = 0.0 

    ! Create predictor from *.pb
    ptr = C_CreatePredictor("simple_graph_tf2.pb")

    ! Register input nodes
    call C_PredictorRegisterInputNode(ptr, "input_a")
    call C_PredictorRegisterInputNode(ptr, "input_b")

    ! Register output nodes
    call C_PredictorRegisterOutputNode(ptr, "result")

    ! Set number of data instances
    call C_PredictorSetDataCount(ptr, 3)

    ! Set the input data
    call C_PredictorSetNodeData(ptr, "input_a", arr_a, 6)
    call C_PredictorSetNodeData(ptr, "input_b", arr_b, 6)

    ! Run the model 
    call C_PredictorRun(ptr)

    ! Get the model output data into arr_c 
    call C_PredictorGetNodeData(ptr, "result", arr_c, 6)

    ! Print the output
    print*, "Calculation Result:", arr_c

    ! Delete predictor when it is not used anymore
    call C_DeletePredictor(ptr)

end program main
\end{lstlisting}

\noindent
\textbf{Entire API list for Fortran}:
\small
\begin{itemize}
  \item \textbf{F01: Loading Model:} 
  \begin{itemize}
    \item \verb|ptr = C_CreatePredictor(file_name)|
    \begin{itemize}
      \item \verb|file_name| -- the Fortran \verb|CHARACTER| array specifying the file name of the PB graph.
    \end{itemize}
    
    To create the \verb|Predictor| object from a \verb|*.pb| format, which return the reference to the \verb|Predictor| object, the return value, \verb|ptr| should be defined as \verb|type(c_ptr)::ptr|.
    \item \verb|ptr = C_CreatePredictor(model_dir, tag)|
    \begin{itemize}
      \item \verb|model_dir| -- the Fortran \verb|CHARACTER| array specifying the directory of the \verb|SavedModel| format (this format itself is a folder).
      \item \verb|tags|   -- the Fortran \verb|CHARACTER| array specifying the tags label within a \verb|SavedModel| format, by default is \verb|serve|.
    \end{itemize}
    
    To create the \verb|Predictor| object from a \verb|SavedModel| format, which return the reference to the \verb|Predictor| object, the return value, \verb|ptr| should be defined as \verb|type(c_ptr)::ptr|.
  \end{itemize}

  \item \textbf{F02: Register Inputs:} 
  \begin{itemize}
    \item \verb|C_PredictorRegisterInputNode(ptr, in_node_name) |
    \begin{itemize}
    \item \verb|ptr| -- the C pointer reference to the created \verb|Predictor| object.
    \item \verb|in_node_name| -- the Fortran \verb|CHARACTER| array specifying the name of the node to be registered as input.
    \end{itemize}
    
    To register input node for feeding data.
  \end{itemize}

  \item \textbf{F03: Register Outputs:} 
  \begin{itemize}
    \item \verb|C_PredictorRegisterOutputNode(ptr, out_node_name) |
    \begin{itemize}
    \item \verb|ptr| -- the C pointer reference to the created \verb|Predictor| object.
    \item \verb|out_node_name| -- the Fortran \verb|CHARACTER| array specifying the name of the node to be registered as output.
    \end{itemize}
    
    To register output node for extracting data.
  \end{itemize}

  \item \textbf{F04: Set Data Counts:} 
  \begin{itemize}
    \item \verb|C_PredictorSetDataCount(ptr, n_data)|
    \begin{itemize}
    \item \verb|ptr| -- the C pointer reference to the created \verb|Predictor| object.
    \item \verb|n_data| -- the Fortran \verb|INTEGER| variable specifying number of data instances.
    \end{itemize}
    
    To set the number of data instances to substitute the unknown dimension of input/output tensors. For example, in the A + B example, the input/output shapes are all [-1, 2], the function could set the unknown -1 into concrete value so that the inner data containers can be created.
  \end{itemize}

  \item \textbf{F05: Set Input Data:} 
  \begin{itemize}

    \item \verb|C_PredictorSetNodeData(ptr, in_node, input_arr , n_element)|
    \begin{itemize} 
    \item \verb|ptr| -- the C pointer reference to the created \verb|Predictor| object.
    \item \verb|in_node| -- the Fortran \verb|CHARACTER| array specifying the name of the input node to be fed with external data.
    \item \verb|input_arr| -- the Fortran numerical array holding the external data, available data type: \verb|INTEGER|, \verb|REAL(4)| and \verb|REAL(8)|.
    \item \verb|n_element| -- the Fortran \verb|INTEGER| variable specifying the number of data elements of the external data array.
    \end{itemize}
    
    To feed the internal input data container registered under \verb|in_node| with external data array. Because the shape information of a Fortran array is lost when passing to a C library, so the total number of elements in the array needs to be specified. If the data type of the internal container and external source are different, this function would automatically cast the datatype to resolve the difference.

    \item \verb|C_PredictorSetNodeDataTranspose(ptr, in_node, input_arr , n_element)|
    \begin{itemize} 
    \item \verb|ptr| -- the C pointer reference to the created \verb|Predictor| object.
    \item \verb|in_node| -- the Fortran \verb|CHARACTER| array specifying the name of the input node to be fed with external data.
    \item \verb|input_arr| -- the Fortran numerical array holding the external data, available data type: \verb|INTEGER|, \verb|REAL(4)| and \verb|REAL(8)|.
    \item \verb|n_element| -- the Fortran \verb|INTEGER| variable specifying the number of data elements of the external data array.
    \end{itemize}
    
    To feed the internal input data container registered under \verb|in_node| with external data array. Because the shape information of a Fortran array is lost when passing to a C library, so the total number of elements in the array needs to be specified. If the data type of the internal container and external source are different, this function would automatically cast the datatype to resolve the difference. The memory layout would be tansposed during the mapping process. 
  \end{itemize}

  \item \textbf{F06: Run Model:} 
  \begin{itemize}
    \item \verb|C_PredictorRun(ptr)|
    \begin{itemize} 
      \item \verb|ptr| -- the C pointer reference to the created \verb|Predictor| object.
    \end{itemize}
    To run the model prediction, the result will be stored in the internal data container holding the model's output.
  \end{itemize}

  \item \textbf{F07: Get Output Data:} 
  \begin{itemize}

    \item \verb|C_PredictorGetNodeData(ptr, out_node, output_arr , n_element)|
    \begin{itemize} 
    \item \verb|ptr| -- the C pointer reference to the created \verb|Predictor| object.
    \item \verb|out_node| -- the Fortran \verb|CHARACTER| array specifying the name of the output node's data to be extracted to external array.
    \item \verb|output_arr| -- the Fortran numerical array holding the external data, available data type: \verb|INTEGER|, \verb|REAL(4)| and \verb|REAL(8)|.
    \item \verb|n_element| -- the Fortran \verb|INTEGER| variable specifying the number of data elements of the external data array.
    \end{itemize}
    
    To extract the data stored in the the internal output data container registered under \verb|out_node| into external data array. Because the shape information of a Fortran array is lost when passing to a C library, so the total number of elements in the array needs to be specified. If the data type of the internal container and external source are different, this function would automatically cast the datatype to resolve the difference.

    \item \verb|C_PredictorGetNodeDataTranspose(ptr, out_node, output_arr , n_element)|
    \begin{itemize} 
    \item \verb|ptr| -- the C pointer reference to the created \verb|Predictor| object.
    \item \verb|out_node| -- the Fortran \verb|CHARACTER| array specifying the name of the output node's data to be extracted to external array.
    \item \verb|output_arr| -- the Fortran numerical array holding the external data, available data type: \verb|INTEGER|, \verb|REAL(4)| and \verb|REAL(8)|.
    \item \verb|n_element| -- the Fortran \verb|INTEGER| variable specifying the number of data elements of the external data array.
    \end{itemize}
    
    To extract the data stored in the the internal output data container registered under \verb|out_node| into external data array. Because the shape information of a Fortran array is lost when passing to a C library, so the total number of elements in the array needs to be specified. If the data type of the internal container and external source are different, this function would automatically cast the datatype to resolve the difference. The memory layout would be tansposed during the mapping process. 
  \end{itemize}

  \item \textbf{F08: Finalize Mode:} 
  \begin{itemize}
    \item \verb|C_DeletePredictor(ptr)|
    \begin{itemize} 
      \item \verb|ptr| -- the C pointer reference to the created \verb|Predictor| object.
    \end{itemize}
    To delete the \verb|Predictor| object pointed by \verb|ptr|, the manual finalization is needed because the resource created by C++ library would not be automatically released by Fortran. 
  \end{itemize}
\end{itemize}

\subsection{OpenFOAM Predictor Module \label{Sec:OpenFOAMPredictor}}
\noindent
The OpenFOAM module specifies all the available setting options in an OpenFOAM dictionary:
\begin{lstlisting}[language=C++,style=mystyle,caption=OpenFOAM dictionary to initialize predictor library, label={lst:OFdict}]
model  // Dictionary name, referenced by the constructor function
{
    readFromPB                yes;               // yes/no for PB/SavedModel    
    modelDirectory            "models/simple_graph_tf2.pb";
//  tags                      "serve";           // only needed for SavedModel     
    copyMethod                Eigen;             // Eigen/Safe for SIMD/element-wise copy    
    layout                    ColMajor;          // RowMajor or ColMajor 
    inputs                    ("input_a" "input_b");    // use space as separator
    outputs                   ("result");               // use space as separator
}
\end{lstlisting}
Therefore, the function call is simplified with just only the constructor function and prediction function (line 31 and 34 in the following codes). But calling the prediction is slightly different since OpenFOAM defines its own fundamental data structures. Two-level container (\texttt{Foam::List}) of the \texttt{Foam::scalarField} references need to be assembled. The outer-level \texttt{Foam::List} corresponds to the number of nodes, whereas the inner-level capacity corresponds to the known dimension of the node shape. For example in the \textbf{A + B} model, two independent nodes are registered as input nodes, so that the outer list capacity of the \texttt{multi\_inputs} is two (line 36). The dimensions of the inner list (e.g., \texttt{multi\_inputs[0]} in line 40) is two because the tensor shape of the input/output node is \texttt{[-1, 2]}. The exact code example can be found below:
\begin{lstlisting}[language=C++,style=mystyle,caption=Minimal example to run A + B model in OpenFOAM, label={lst:AplusB_OF}]
#include "scalarField.H"      // Header from OpenFOAM
#include "TF_OF_Predictor.H"  // Header for TF_Predictor

int main() {

    // External OF fields ...

    // create input field A (columns):
    scalarField input_field_a_col1(3, scalar(0.0));
    input_field_a_col1[0] = 0.0;
    input_field_a_col1[1] = 2.2;
    input_field_a_col1[2] = 4.4;
    scalarField input_field_a_col2(3, scalar(0.0));
    input_field_a_col2[0] = 1.1;
    input_field_a_col2[1] = 3.3;
    input_field_a_col2[2] = 5.5;


    // create input field B (columns):
    scalarField input_field_b_col1(3, scalar(0.0));
    input_field_b_col1[0] = 5.0;
    input_field_b_col1[1] = 3.0;
    input_field_b_col1[2] = 1.0;
    scalarField input_field_b_col2(3, scalar(0.0));
    input_field_b_col2[0] = 4.0;
    input_field_b_col2[1] = 2.0;
    input_field_b_col2[2] = 0.0;

    // create output field C (columns)
    scalarField output_field_c_col1(3, scalar(0.0)); // output field for pb model column1
    scalarField output_field_c_col2(3, scalar(0.0)); // output field for pb model column2

 
    // Create the entry list for OF fields:
    // PB format model:
    Foam::List<Foam::List<scalarField*>> multi_inputs(2);
    Foam::List<Foam::List<scalarField*>> multi_outputs(1);

    // For input_a:
    multi_inputs[0].append(&input_field_a_col1);
    multi_inputs[0].append(&input_field_a_col2);

    // For input_b:
    multi_inputs[1].append(&input_field_b_col1);
    multi_inputs[1].append(&input_field_b_col2);

    // For output_c:
    multi_outputs[0].append(&output_field_c_col1);
    multi_outputs[0].append(&output_field_c_col2);


    Info << "Creating TF_OF_Predictor\n" << endl;
    TF_OF_Predictor pd = TF_OF_Predictor("constant/TF_Predictor_Dict", "model_pb_2D");


    Info << "Running the models \n" << endl;
    pd.predict(multi_inputs, multi_outputs);

    Info << "Print the running results\n" << endl;
    Info << "Results of column 1: \n"
         << input_field_a_col1 << " + " << input_field_b_col1 << " = " << output_field_c_col1 << endl;

    Info << "Results of column 2: \n"
         << input_field_a_col2 << " + " << input_field_b_col2 << " = " << output_field_c_col2 << endl;

    return 0;
}
\end{lstlisting}

\noindent
\textbf{Entire API list for OpenFOAM Module}:
\small
\begin{itemize}
  \item \textbf{OF01: Model Initialization:} 
  \begin{itemize}
    \item \verb|TF_OF_Predictor()|
    
    To create the \verb|TF_OF_Predictor| object from the OpenFOAM dictionary stored in default directory. The default file path of the dictionary is \verb|constant/TF_Predictor_Dict|, and the settings are under the default dictionary name, \verb|model|.
    \item \verb|TF_OF_Predictor(std::string Model_name)|
    \begin{itemize}
      \item \verb|Model_name| -- the name of the dictionary the model will be launched under its setting options.
    \end{itemize}
    
    To create the \verb|TF_OF_Predictor| object from the OpenFOAM dictionary stored in default directory. The default file path of the dictionary is \verb|constant/TF_Predictor_Dict|, and the settings are under the dictionary name, \verb|Model_name|.

    \item \verb|TF_OF_Predictor(std::string Dict_dir, std::string Model_name)|
    \begin{itemize}
      \item \verb|Dict_dir| -- the OpenFOAM dictionary file holding the dictionaries of the settings of the \verb|TF_OF_Predictor|.
      \item \verb|Model_name| -- the name of the dictionary the model will be launched under its setting options.
    \end{itemize}
    
    To create the \verb|TF_OF_Predictor| object from the OpenFOAM dictionary stored in \verb|Dict_dir|, and the settings are under the dictionary name, \verb|Model_name|.
  \end{itemize}

  \item \textbf{OF02: Model Prediction:} 
  \begin{itemize}
    \item \verb|predict(Foam::List<Foam::scalarField*>& inputs,|\\
          \verb|        Foam::List<Foam::scalarField*>& outputs)|
    \begin{itemize}
    \item \verb|inputs| -- the reference container holding the references of input \verb|scalarField| for a single-input-output model.
    \item \verb|outputs| -- the reference container holding the references of output \verb|scalarField| for a single-input-output model.
    \end{itemize}
    
    To perform prediction of a single-input-output model, which means the number of the input nodes and output nodes are both \textbf{one}. In this case, the capacity of the reference container corresponds to the known dimension of the node shape. For example, if the node shape is \texttt{[-1, 2, 3]}, then the reference container needs to have $2 \times 3 = 6$ references of \verb|scalarField|.

    \item \verb|predict(Foam::List<Foam::List<Foam::scalarField*>> & multi_inputs,|\\
          \verb|        Foam::List<Foam::List<Foam::scalarField*>> & multi_outputs)|
    \begin{itemize}
    \item \verb|inputs| -- the two-level reference container holding the references of input \verb|scalarField| for a multiple-input-output model.
    \item \verb|outputs| -- the two-level reference container holding the references of output \verb|scalarField| for a multiple-input-output model.
    \end{itemize}

    To perform prediction of a multiple-input-output model, which means model has more than one input nodes or output nodes. In this case, the outer-level capacity of the two-level reference container equals to the number of nodes (e.g., the \verb|multi_inputs| in the example), and the inner-level capacity corresponds to the known dimension of the node shape. For example, if the node shape is \texttt{[-1, 2, 3]}, then the reference container needs to have $2 \times 3 = 6$ references of \verb|scalarField|.

    \item \verb|predict(Foam::List<Foam::volScalarField*>& inputs,|\\
          \verb|        Foam::List<Foam::volScalarField*>& outputs)|
    \begin{itemize}
    \item \verb|inputs| -- the reference container holding the references of input \verb|volScalarField| for a single-input-output model.
    \item \verb|outputs| -- the reference container holding the references of output \verb|volScalarField| for a single-input-output model.
    \end{itemize}
    
    To perform prediction of a single-input-output model, which means the number of the input nodes and output nodes are both \textbf{one}. In this case, the capacity of the reference container corresponds to the known dimension of the node shape. For example, if the node shape is \texttt{[-1, 2, 3]}, then the reference container needs to have $2 \times 3 = 6$ references of \verb|volScalarField|.

    \item \verb|predict(Foam::List<Foam::List<Foam::volScalarField*>> & multi_inputs,|\\
          \verb|        Foam::List<Foam::List<Foam::volScalarField*>> & multi_outputs)|
    \begin{itemize}
    \item \verb|inputs| -- the two-level reference container holding the references of input \verb|volScalarField| for a multiple-input-output model.
    \item \verb|outputs| -- the two-level reference container holding the references of output \verb|volScalarField| for a multiple-input-output model.
    \end{itemize}

    To perform prediction of a multiple-input-output model, which means model has more than one input nodes or output nodes. In this case, the outer-level capacity of the two-level reference container equals to the number of nodes (e.g., the \verb|multi_inputs| in the example), and the inner-level capacity corresponds to the known dimension of the node shape. For example, if the node shape is \texttt{[-1, 2, 3]}, then the reference container needs to have $2 \times 3 = 6$ references of \verb|volScalarField|.
  \end{itemize}

\end{itemize}

\subsection{Fortran Module Template (Experience in Modifying CFL3D)\label{Sec:FortranPredictorModule}}
The \href{https://github.com/Weishuo93/NN_Pred/blob/master/CFL3D-Extension/Solvers/Code_Template/ML_Template_Module.F90}{module template} to integrate the predictor into a Fortran CFD program can be found in the public repository. The design of this module allows it to be used as an absolute external plug-in component, which means the module maintains all the user-defined variables (e.g., predictor instances, additional field) by itself, and only interacts with the program's originally allocated arrays. In this way, no extra memory is allocated in the main program, and the risk of memory leak caused by mutual contamination between the legacy-code-created global variables (normally in F77 format) and newly created objects can be minimized to the greatest extent.

The structural template of the module can be viewed in the repository: \href{https://github.com/Weishuo93/NN_Pred/blob/master/CFL3D-Extension/Solvers/Code_Template/ML_Template_Module.F90}{\texttt{ML\_Template\_Module.F90}}. Basically, all the managements of the ML-associated variables (e.g., predictor objects and data arrays) are governed by the module itself. whereas the subroutine for updating field is in charge of interacting with the main program. Users only need to define their needed arrays in the definition of \texttt{pd\_pack}, the initialization/finalization functions for such arrays, and the update-field subroutine by themselves. 

A reference of the read-in text file for predictor settings is also shown below as a reference (Listing~\ref{lst:FortranSetting}), the initialization and the format of this setting file also can be modified according to the user's demand.

\begin{lstlisting}[language=bash,style=mystyle,caption=Setting file for Fortran predictor module, label={lst:FortranSetting}]
# Whether it is from PB (0) or from SavedModel (1) format? !n_ml_options;
0
# This is model dir
ML_inputs/output_graph.pb
# This is the tag string for SavedModel(1) format, will not be used if n_ml_options = 0
serve
# This is input node name
Lamda_in/Placeholder
# This is output node name
layer_out/Wx_plus_b/Add
\end{lstlisting}




\bibliographystyle{elsarticle-num}
\bibliography{TF_predictor}







\end{document}